\documentclass[a4paper,twocolumn,11pt,unpublished]{quantumarticle}
\pdfoutput=1
\usepackage[caption=false]{subfig} 	
\usepackage{graphicx}  				
\graphicspath{{figure/}}
\usepackage{mathtools}				
\usepackage{amssymb}   				
\usepackage{amsthm}    				
\usepackage{acronym}				
\usepackage{algorithmicx}			
\usepackage[noend]{algpseudocode}	
\usepackage{tabularx}				
\usepackage{etoolbox}				
\usepackage{placeins}				
\usepackage{tcolorbox}				
\usepackage{paralist}				
\usepackage{enumitem}				
\usepackage{listings}				
\usepackage{verbatim}				
\usepackage[numbers,sort&compress]{natbib}
\usepackage{hyperref}
\usepackage{xcolor}

\newtheorem{lemma}{Lemma}

\newtheorem{proposition}{Proposition}
\newtheorem{theorem}{Theorem}

\makeatletter
\renewcommand{\maketag@@@}[1]{\hbox{\m@th\normalsize\normalfont#1}}%
\makeatother

\usepackage{xcolor}
\definecolor{Jblue}{HTML}{0066CC}
\definecolor{Jgrey}{HTML}{8A8888}
\definecolor{Jred}{HTML}{D70040}
\definecolor{Jgreen}{HTML}{666633}

\newif\ifcmnt
\cmnttrue

\ifcmnt
	\newenvironment{cmng}{\begingroup\color{Jgrey}}{\endgroup}
	\newenvironment{cmnb}{\begingroup\color{Jblue}}{\endgroup}
	\newenvironment{cmnr}{\begingroup\color{Jred}}{\endgroup}
\else

\fi

\usepackage{soul}
\setulcolor{red}

\allowdisplaybreaks

\begin{document}

\title{Quantum Ring States}

\author{Shang-Jen Su}
\affiliation{Georgia Institute of Technology, Atlanta, Georgia 30332, USA}

\author{Shi-Yuan Wang}
\affiliation{Qualcomm Technologies, Inc., San Diego, California 92121, USA}
\thanks{S.-Y. Wang contributed to the present work while pursuing a Ph.D. degree at the Georgia Institute of Technology.}

\author{Tuna Erdo\u{g}an}
\affiliation{Georgia Institute of Technology, Atlanta, Georgia 30332, USA}

\author{Zheshen Zhang}
\affiliation{University of Michigan, Ann Arbor, Michigan 48109, USA}

\author{Matthieu R. Bloch}
\affiliation{Georgia Institute of Technology, Atlanta, Georgia 30332, USA}

\begin{abstract}
Quantum ring states are non-Gaussian mixed states generated by uniformly modulating the phase of one arm of a bipartite Gaussian quantum resource and transmitting the modulated arm through a lossy thermal bosonic channel.
For resources such as two-mode squeezed vacuum (TMSV) states and split coherent states, the continuous phase modulation produces classical states that are diagonal in the Fock basis and fully characterized by photon-number distributions involving hypergeometric functions.
We precisely characterize how well states obtained with a finite $m$-ary phase-shift keying (PSK) modulation approximate quantum ring states.
We also leverage bounds on hypergeometric functions to develop closed-form and surprisingly tight bounds for information-theoretic quantities involving quantum ring states, such as the von Neumann entropy and the associated Holevo information.
We demonstrate the usefulness of quantum ring states by revisiting several canonical problems and deriving new results including: 1)~closed-form achievable communication rates with PSK modulation for lossy thermal bosonic channels over a broad range of channel parameters; 2)~improved achievable covert throughputs for one-way and round-trip lossy thermal bosonic channels.
\end{abstract}

\maketitle

\acrodef{ACDIS}[ACDIS]{Adaptive Communication Decision and Information Systems}
\acrodef{AEP}{Asymptotic Equipartition Property}
\acrodef{AoA}{Angle of Arrival}
\acrodef{AWGN}{Additive White Gaussian Noise}
\acrodef{AVC}[AVC]{Arbitrarily Varying Channel}
\acrodef{BER}{Bit-Error-Rate}
\acrodef{BEC}{Binary Erasure Channel}
\acrodef{BSC}{Binary Symmetric Channel}
\acrodef{BICM}[BICM]{Bit-Interleaved Coded-Modulation}
\acrodef{CDF}[CDF]{cumulative distribution function}
\acrodef{CGF}[CGF]{Cumulant Generating Function}
\acrodef{CLT}[CLT]{Central Limit Theorem}
\acrodef{CSI}[CSI]{Channel State Information}
\acrodef{DMC}[DMC]{Discrete Memoryless Channel}
\acrodef{DMS}[DMS]{Discrete Memoryless Source}
\acrodef{ERM}[ERM]{Empirical Risk Minimization}
\acrodef{FER}[FER]{Frame Error Rate}
\acrodef{ICA}[ICA]{Independent Component Analysis}
\acrodef{IoT}[IoT]{Internet of Things}
\acrodef{KKT}[KKT]{Karush-Kuhn Tucker}
\acrodef{LASSO}[LASSO]{Least Absolute Shrinkage and Selection Operator}
\acrodef{LPD}[LPD]{Low Probability of Detection}
\acrodef{LDPC}[LDPC]{Low-Density Parity-Check}
\acrodef{LLMS}[LLMS]{Linear Least Mean Square}
\acrodef{LMS}[LMS]{Least Mean Square}
\acrodef{MAC}[MAC]{multiple-access channel}
\acrodef{MGF}[MGF]{Moment Generating Function}
\acrodef{MLC}[MLC]{Multi-Level Coding}
\acrodef{MLE}[MLE]{Maximum Likelihood Estimate}
\acrodef{MIMO}[MIMO]{Multiple-Input Multiple-Output}
\acrodef{MISO}{Multiple-Input Single-Output}
\acrodef{MSD}[MSD]{Multi-Stage Decoding}
\acrodef{MMSE}[MMSE]{Minimum Mean-Square Error}
\acrodef{PAC}[PAC]{Probably Approximately Correct}
\acrodef{PCA}[PCA]{Principal Component Analysis}
\acrodef{PDF}[PDF]{Probability Density Function}
\acrodef{PMF}[PMF]{Probability Mass Function}
\acrodef{POVM}[POVM]{Positive Operator-Valued Measure}
\acrodef{PVM}[PVM]{Projection-Valued Measure}
\acrodef{PPM}[PPM]{Pulse Position Modulation}
\acrodef{PSD}{Power Spectral Density}
\acrodef{QKD}{Quantum Key Distribution}
\acrodef{ROC}{Receiver Operating Characteristic}
\acrodef{CVQKD}{Continuous-Variable \ac{QKD}}
\acrodef{QPSK}{Quadrature Phase-Shift Keying}
\acrodef{RV}{random variable}
\acrodef{SFG}{Sum Frequency Generation}
\acrodef{SIMO}{Single-Input Multiple-Output}
\acrodef{SNR}{Signal-to-Noise Ratio}
\acrodef{SVM}[SVM]{Support Vector Machine}
\acrodef{TMSV}[TMSV]{two-mode squeezed vacuum}
\acrodef{ASE}[ASE]{Amplified Spontaneous Emission}
\acrodef{TPCP}{Trace-Preserving Completely-Positive}
\acrodef{wrt}[w.r.t.]{with respect to}
\acrodef{WSS}{Wide Sense Stationary}

\acrodef{PSK}{phase-shift keying}
\acrodef{QPSK}{quadrature phase-shift keying}
\acrodef{BPSK}{binary phase-shift keying}
\acrodef{i.i.d.}[i.i.d.]{independent and identically distributed}
\acrodef{cq}[cq]{classical-quantum}
\acrodef{HSW}[HSW]{Holevo-Schumacher-Westmoreland}


\newcommand{\matA}{\mathbf{A}}
\newcommand{\bfA}{\mathbf{A}}
\newcommand{\calA}{\mathcal{A}}
\newcommand{\bbA}{\mathbb{A}}
\newcommand{\bfa}{\mathbf{a}}
\newcommand{\veca}{\mathbf{a}}

\newcommand{\matB}{\mathbf{B}}
\newcommand{\bfB}{\mathbf{B}}
\newcommand{\calB}{\mathcal{B}}
\newcommand{\bbB}{\mathbb{B}}
\newcommand{\bfb}{\mathbf{b}}
\newcommand{\vecb}{\mathbf{b}}

\newcommand{\matC}{\mathbf{C}}
\newcommand{\bfC}{\mathbf{C}}
\newcommand{\calC}{\mathcal{C}}
\newcommand{\bbC}{\mathbb{C}}
\newcommand{\bfc}{\mathbf{c}}
\newcommand{\vecc}{\mathbf{c}}

\newcommand{\matD}{\mathbf{D}}
\newcommand{\bfD}{\mathbf{D}}
\newcommand{\calD}{\mathcal{D}}
\newcommand{\bbD}{\mathbb{D}}
\newcommand{\bfd}{\mathbf{d}}
\newcommand{\vecd}{\mathbf{d}}

\newcommand{\matE}{\mathbf{E}}
\newcommand{\bfE}{\mathbf{E}}
\newcommand{\calE}{\mathcal{E}}
\newcommand{\bbE}{\mathbb{E}}
\newcommand{\bfe}{\mathbf{e}}
\newcommand{\vece}{\mathbf{e}}

\newcommand{\matF}{\mathbf{F}}
\newcommand{\bfF}{\mathbf{F}}
\newcommand{\calF}{\mathcal{F}}
\newcommand{\bbF}{\mathbb{F}}
\newcommand{\bff}{\mathbf{f}}
\newcommand{\vecf}{\mathbf{f}}

\newcommand{\matG}{\mathbf{G}}
\newcommand{\bfG}{\mathbf{G}}
\newcommand{\calG}{\mathcal{G}}
\newcommand{\bbG}{\mathbb{G}}
\newcommand{\bfg}{\mathbf{g}}
\newcommand{\vecg}{\mathbf{g}}

\newcommand{\matH}{\mathbf{H}}
\newcommand{\bfH}{\mathbf{H}}
\newcommand{\calH}{\mathcal{H}}
\newcommand{\bbH}{\mathbb{H}}
\newcommand{\bfh}{\mathbf{h}}
\newcommand{\vech}{\mathbf{h}}

\newcommand{\matI}{\mathbf{I}}
\newcommand{\bfI}{\mathbf{I}}
\newcommand{\calI}{\mathcal{I}}
\newcommand{\bbI}{\mathbb{I}}
\newcommand{\bfi}{\mathbf{i}}
\newcommand{\veci}{\mathbf{i}}

\newcommand{\matJ}{\mathbf{J}}
\newcommand{\bfJ}{\mathbf{J}}
\newcommand{\calJ}{\mathcal{J}}
\newcommand{\bbJ}{\mathbb{J}}
\newcommand{\bfj}{\mathbf{j}}
\newcommand{\vecj}{\mathbf{j}}

\newcommand{\matK}{\mathbf{K}}
\newcommand{\bfK}{\mathbf{K}}
\newcommand{\calK}{\mathcal{K}}
\newcommand{\bbK}{\mathbb{K}}
\newcommand{\bfk}{\mathbf{k}}
\newcommand{\veck}{\mathbf{k}}

\newcommand{\matL}{\mathbf{L}}
\newcommand{\bfL}{\mathbf{L}}
\newcommand{\calL}{\mathcal{L}}
\newcommand{\bbL}{\mathbb{L}}
\newcommand{\bfl}{\mathbf{l}}
\newcommand{\vecl}{\mathbf{l}}

\newcommand{\matM}{\mathbf{M}}
\newcommand{\bfM}{\mathbf{M}}
\newcommand{\calM}{\mathcal{M}}
\newcommand{\bbM}{\mathbb{M}}
\newcommand{\bfm}{\mathbf{m}}
\newcommand{\vecm}{\mathbf{m}}

\newcommand{\matN}{\mathbf{N}}
\newcommand{\bfN}{\mathbf{N}}
\newcommand{\calN}{\mathcal{N}}
\newcommand{\bbN}{\mathbb{N}}
\newcommand{\bfn}{\mathbf{n}}
\newcommand{\vecn}{\mathbf{n}}

\newcommand{\matO}{\mathbf{O}}
\newcommand{\bfO}{\mathbf{O}}
\newcommand{\calO}{\mathcal{O}}
\newcommand{\bbO}{\mathbb{O}}
\newcommand{\bfo}{\mathbf{o}}
\newcommand{\veco}{\mathbf{o}}

\newcommand{\matP}{\mathbf{P}}
\newcommand{\bfP}{\mathbf{P}}
\newcommand{\calP}{\mathcal{P}}
\newcommand{\bbP}{\mathbb{P}}
\newcommand{\bfp}{\mathbf{p}}
\newcommand{\vecp}{\mathbf{p}}

\newcommand{\matQ}{\mathbf{Q}}
\newcommand{\bfQ}{\mathbf{Q}}
\newcommand{\calQ}{\mathcal{Q}}
\newcommand{\bbQ}{\mathbb{Q}}
\newcommand{\bfq}{\mathbf{q}}
\newcommand{\vecq}{\mathbf{q}}

\newcommand{\matR}{\mathbf{R}}
\newcommand{\bfR}{\mathbf{R}}
\newcommand{\calR}{\mathcal{R}}
\newcommand{\bbR}{\mathbb{R}}
\newcommand{\bfr}{\mathbf{r}}
\newcommand{\vecr}{\mathbf{r}}

\newcommand{\matS}{\mathbf{S}}
\newcommand{\bfS}{\mathbf{S}}
\newcommand{\calS}{\mathcal{S}}
\newcommand{\bbS}{\mathbb{S}}
\newcommand{\bfs}{\mathbf{s}}
\newcommand{\vecs}{\mathbf{s}}

\newcommand{\matT}{\mathbf{T}}
\newcommand{\bfT}{\mathbf{T}}
\newcommand{\calT}{\mathcal{T}}
\newcommand{\bbT}{\mathbb{T}}
\newcommand{\bft}{\mathbf{t}}
\newcommand{\vect}{\mathbf{t}}

\newcommand{\matU}{\mathbf{U}}
\newcommand{\bfU}{\mathbf{U}}
\newcommand{\calU}{\mathcal{U}}
\newcommand{\bbU}{\mathbb{U}}
\newcommand{\bfu}{\mathbf{u}}
\newcommand{\vecu}{\mathbf{u}}

\newcommand{\matV}{\mathbf{V}}
\newcommand{\bfV}{\mathbf{V}}
\newcommand{\calV}{\mathcal{V}}
\newcommand{\bbV}{\mathbb{V}}
\newcommand{\bfv}{\mathbf{v}}
\newcommand{\vecv}{\mathbf{v}}

\newcommand{\matW}{\mathbf{W}}
\newcommand{\bfW}{\mathbf{W}}
\newcommand{\calW}{\mathcal{W}}
\newcommand{\bbW}{\mathbb{W}}
\newcommand{\bfw}{\mathbf{w}}
\newcommand{\vecw}{\mathbf{w}}

\newcommand{\matX}{\mathbf{X}}
\newcommand{\bfX}{\mathbf{X}}
\newcommand{\calX}{\mathcal{X}}
\newcommand{\bbX}{\mathbb{X}}
\newcommand{\bfx}{\mathbf{x}}
\newcommand{\vecx}{\mathbf{x}}

\newcommand{\matY}{\mathbf{Y}}
\newcommand{\bfY}{\mathbf{Y}}
\newcommand{\calY}{\mathcal{Y}}
\newcommand{\bbY}{\mathbb{Y}}
\newcommand{\bfy}{\mathbf{y}}
\newcommand{\vecy}{\mathbf{y}}

\newcommand{\matZ}{\mathbf{Z}}
\newcommand{\bfZ}{\mathbf{Z}}
\newcommand{\calZ}{\mathcal{Z}}
\newcommand{\bbZ}{\mathbb{Z}}
\newcommand{\bfz}{\mathbf{z}}
\newcommand{\vecz}{\mathbf{z}}

\newcommand{\bfalpha}{\boldsymbol{\alpha}}
\newcommand{\vecalpha}{\boldsymbol{\alpha}}

\newcommand{\bfbeta}{\boldsymbol{\beta}}
\newcommand{\vecbeta}{\boldsymbol{\beta}}

\newcommand{\bfgamma}{\boldsymbol{\gamma}}
\newcommand{\vecgamma}{\boldsymbol{\gamma}}

\newcommand{\bfdelta}{\boldsymbol{\delta}}
\newcommand{\vecdelta}{\boldsymbol{\delta}}

\newcommand{\bfepsilon}{\boldsymbol{\epsilon}}
\newcommand{\vecepsilon}{\boldsymbol{\epsilon}}

\newcommand{\bfzeta}{\boldsymbol{\zeta}}
\newcommand{\veczeta}{\boldsymbol{\zeta}}

\newcommand{\bfeta}{\boldsymbol{\eta}}
\newcommand{\veceta}{\boldsymbol{\eta}}

\newcommand{\bftheta}{\boldsymbol{\theta}}
\newcommand{\vectheta}{\boldsymbol{\theta}}

\newcommand{\bfiota}{\boldsymbol{\iota}}
\newcommand{\veciota}{\boldsymbol{\iota}}

\newcommand{\bfkappa}{\boldsymbol{\kappa}}
\newcommand{\veckappa}{\boldsymbol{\kappa}}

\newcommand{\bflamda}{\boldsymbol{\lamda}}
\newcommand{\veclamda}{\boldsymbol{\lamda}}

\newcommand{\bfmu}{\boldsymbol{\mu}}
\newcommand{\vecmu}{\boldsymbol{\mu}}

\newcommand{\bfnu}{\boldsymbol{\nu}}
\newcommand{\vecnu}{\boldsymbol{\nu}}

\newcommand{\bfxi}{\boldsymbol{\xi}}
\newcommand{\vecxi}{\boldsymbol{\xi}}

\newcommand{\bfomicron}{\boldsymbol{\omicron}}
\newcommand{\vecomicron}{\boldsymbol{\omicron}}

\newcommand{\bfpi}{\boldsymbol{\pi}}
\newcommand{\vecpi}{\boldsymbol{\pi}}

\newcommand{\bfrho}{\boldsymbol{\rho}}
\newcommand{\vecrho}{\boldsymbol{\rho}}

\newcommand{\bfsigma}{\boldsymbol{\sigma}}
\newcommand{\vecsigma}{\boldsymbol{\sigma}}

\newcommand{\bftau}{\boldsymbol{\tau}}
\newcommand{\vectau}{\boldsymbol{\tau}}

\newcommand{\bfupsilon}{\boldsymbol{\upsilon}}
\newcommand{\vecupsilon}{\boldsymbol{\upsilon}}

\newcommand{\bfphi}{\boldsymbol{\phi}}
\newcommand{\vecphi}{\boldsymbol{\phi}}

\newcommand{\bfchi}{\boldsymbol{\chi}}
\newcommand{\vecchi}{\boldsymbol{\chi}}

\newcommand{\bfpsi}{\boldsymbol{\psi}}
\newcommand{\vecpsi}{\boldsymbol{\psi}}

\newcommand{\bfomega}{\boldsymbol{\omega}}
\newcommand{\vecomega}{\boldsymbol{\omega}}

\newcommand{\bfAlpha}{\boldsymbol{\Alpha}}
\newcommand{\matAlpha}{\boldsymbol{\Alpha}}

\newcommand{\bfBeta}{\boldsymbol{\Beta}}
\newcommand{\matBeta}{\boldsymbol{\Beta}}

\newcommand{\bfGamma}{\boldsymbol{\Gamma}}
\newcommand{\matGamma}{\boldsymbol{\Gamma}}

\newcommand{\bfDelta}{\boldsymbol{\Delta}}
\newcommand{\matDelta}{\boldsymbol{\Delta}}

\newcommand{\bfEpsilon}{\boldsymbol{\Epsilon}}
\newcommand{\matEpsilon}{\boldsymbol{\Epsilon}}

\newcommand{\bfZeta}{\boldsymbol{\Zeta}}
\newcommand{\matZeta}{\boldsymbol{\Zeta}}

\newcommand{\bfEta}{\boldsymbol{\Eta}}
\newcommand{\matEta}{\boldsymbol{\Eta}}

\newcommand{\bfTheta}{\boldsymbol{\Theta}}
\newcommand{\matTheta}{\boldsymbol{\Theta}}

\newcommand{\bfIota}{\boldsymbol{\Iota}}
\newcommand{\matIota}{\boldsymbol{\Iota}}

\newcommand{\bfKappa}{\boldsymbol{\Kappa}}
\newcommand{\matKappa}{\boldsymbol{\Kappa}}

\newcommand{\bfLamda}{\boldsymbol{\Lamda}}
\newcommand{\matLamda}{\boldsymbol{\Lamda}}

\newcommand{\bfMu}{\boldsymbol{\Mu}}
\newcommand{\matMu}{\boldsymbol{\Mu}}

\newcommand{\bfNu}{\boldsymbol{\Nu}}
\newcommand{\matNu}{\boldsymbol{\Nu}}

\newcommand{\bfXi}{\boldsymbol{\Xi}}
\newcommand{\matXi}{\boldsymbol{\Xi}}

\newcommand{\bfOmicron}{\boldsymbol{\Omicron}}
\newcommand{\matOmicron}{\boldsymbol{\Omicron}}

\newcommand{\bfPi}{\boldsymbol{\Pi}}
\newcommand{\matPi}{\boldsymbol{\Pi}}

\newcommand{\bfRho}{\boldsymbol{\Rho}}
\newcommand{\matRho}{\boldsymbol{\Rho}}

\newcommand{\bfSigma}{\boldsymbol{\Sigma}}
\newcommand{\matSigma}{\boldsymbol{\Sigma}}

\newcommand{\bfTau}{\boldsymbol{\Tau}}
\newcommand{\matTau}{\boldsymbol{\Tau}}

\newcommand{\bfUpsilon}{\boldsymbol{\Upsilon}}
\newcommand{\matUpsilon}{\boldsymbol{\Upsilon}}

\newcommand{\bfPhi}{\boldsymbol{\Phi}}
\newcommand{\matPhi}{\boldsymbol{\Phi}}

\newcommand{\bfChi}{\boldsymbol{\Chi}}
\newcommand{\matChi}{\boldsymbol{\Chi}}

\newcommand{\bfPsi}{\boldsymbol{\Psi}}
\newcommand{\matPsi}{\boldsymbol{\Psi}}

\newcommand{\bfOmega}{\boldsymbol{\Omega}}
\newcommand{\matOmega}{\boldsymbol{\Omega}}

\newcommand{\bfzero}{\boldsymbol{0}}

\renewcommand{\leq}{\leqslant}
\renewcommand{\geq}{\geqslant}

\newcommand{\eqdef}{\triangleq}
\newcommand{\argmin}{\mathop{\text{argmin}}}
\newcommand{\argmax}{\mathop{\text{argmax}}}

\newcommand{\abs}[1]{{\left|{#1}\right|}}
\newcommand{\norm}[2][]{{\left\Vert{#2}\right\Vert}_{#1}}
\newcommand{\set}[1]{{\{#1\}}}
\newcommand{\intseq}[2]{[{#1};{#2}]}
\newcommand{\bra}[1]{{\left\langle{#1}\right\vert}}
\newcommand{\ket}[1]{{\left\vert{#1}\right\rangle}}
\newcommand{\braket}[2]{\langle #1 | #2 \rangle}
\newcommand{\kb}[1]{{\left\vert{#1}\right\rangle\left\langle{#1}\right\vert}}
\newcommand{\tr}[2][]{\ensuremath{\text{\textnormal{tr}}_{#1}\left(#2\right)}}  
\renewcommand{\det}[1]{{\left|{#1}\right|}}
\newcommand{\Fid}[1]{\ensuremath{\text{\textnormal{F}}}\left(#1\right)}

\newcommand{\D}[2]{{\mathbb{D}}\!\left(#1\,\middle\Vert\,#2\right)}

\newcommand{\rD}[3][\alpha]{{\mathbb{D}_{#1}}\!\left(#2\,\middle\Vert\,#3\right)}

\newcommand{\tsrD}[3][\alpha]{{\widetilde{\mathbb{D}}_{#1}^{\epsilon}}\!\left(#2\,\middle\Vert\,#3\right)}

\newcommand{\trD}[3][\alpha]{{\widetilde{\mathbb{D}}_{#1}}\!\left(#2\,\middle\Vert\,#3\right)}

\newcommand{\V}[2]{{\mathbb{V}}\!\left(#1\,\middle\Vert\,#2\right)}

\newcommand{\rV}[3][\alpha]{{\mathbb{V}_{#1}}\!\left(#2\,\middle\Vert\,#3\right)}

\newcommand{\I}[2]{{{\mathbb{I}}\!\left(#1;#2\right)}}

\newcommand{\vonH}[1]{{\mathbb{H}}\!\left(#1\right)}

\newcommand{\rH}[2][\alpha]{\mathbb{H}_{#1}\left(#2\right)}

\newcommand{\srH}[2][\alpha]{\mathbb{H}_{#1}^{\epsilon}\left(#2\right)}

\newcommand{\smH}[2][\epsilon]{\mathbb{H}_{\text{\textnormal{min}}}^{#1}\left(#2\right)}

\newcommand{\sMH}[2][\epsilon]{\mathbb{H}_{\text{\textnormal{max}}}^{#1}\left(#2\right)}

\newcommand{\Pdist}[1]{\ensuremath{\text{\textnormal{P}}}\left(#1\right)}

\newcommand{\ball}[2][]{\calB^{#1}\left(#2\right)}

\section{Introduction}
\label{sec:intro}
Bosonic channels provide a fundamental model for optical quantum information processing, encompassing communication and sensing tasks. In particular, bosonic channels provide an accurate quantum description of electromagnetic wave propagation in linear media, enabling the study of applications as diverse as communication over optical and microwave channels, storage in optical memories, and radar systems~\cite{Holevo2001EvaluatingCapacities,Giovannetti2004ClassicalCapacity,Zhao2025QuantumIllumination}. One major success of quantum information theory has been the closed-form characterization of the information-theoretic limits of information processing tasks involving bosonic channels, including classical and entanglement-assisted communication capacities~\cite{Holevo2001EvaluatingCapacities,Bennett2002EntanglementassistedCapacity,Giovannetti2014UltimateClassical} and the quantum Chernoff bound for sensing~\cite{Audenaert2007DiscriminatingStates,Pirandola2008ComputableBounds}, thereby providing useful analytical insights.

We develop further analytical insights by leveraging what we call quantum ring states. A quantum ring state is constructed by uniformly phase modulating one arm of a bipartite Gaussian quantum resource, such as the signal arm of a~\ac{TMSV} state~\cite{Shi2020PracticalRoute} or coherent states~\cite{Tanggara2024QuantumoptimalInformation}, and transmitting the modulated arm through a lossy thermal bosonic channel. For \ac{TMSV} states, it is known from~\cite[(B5)]{Shi2020PracticalRoute} that the resulting mixed state is non-Gaussian, yet diagonal in the Fock basis and fully characterized by a probability mass function involving a hypergeometric function. Perhaps surprisingly, by appropriately bounding the hypergeometric functions, we show that several information-theoretic quantities involving quantum ring states admit simple closed-form expressions.

Our first application of quantum ring states is to derive closed-form achievable communication rates with \ac{PSK} modulation over lossy thermal bosonic channels. Our rates remain valid over a broad range of channel parameters, without relying on commonly used simplifying assumptions such as extremely high noise, low signal power, or extreme transmissivities~\cite{Shi2020PracticalRoute,DiCandia2021TwowayCovert,Cox2023TransceiverDesigns}. Our approach also confirms the near-optimality of phase modulation for achieving capacity, as already observed numerically~\cite{Cox2023TransceiverDesigns}.

Our second application of quantum ring states is to quantum covert communication, in which a sender attempts to transmit messages to a receiver while preventing an eavesdropper from detecting the transmission~\cite{Gagatsos2020CovertCapacity,DiCandia2021TwowayCovert,Wang2022CharacterizationCovert,Wang2024ResourceefficientEntanglementassisted}. We consider both one-way and round-trip covert communication protocols, using either \ac{TMSV} states or coherent states as resources. Our approach allows us to develop several insights. For one-way \ac{TMSV}-assisted covert communication, we show that the covert capacity is achievable with exponentially fewer entangled pairs than in prior work~\cite{Gagatsos2020CovertCapacity}. For round-trip \ac{TMSV}-assisted covert communication, we show that the covert throughput scales as $O \left( \sqrt{n} \log n \right)$ with the number of channel uses $n$, compared to $O(\sqrt{n})$ in previous work~\cite{DiCandia2021TwowayCovert,Fessatidis2026PracticalTwoWay}.

\section{Quantum Ring States}
\label{sec:ring_state}
A lossy thermal bosonic channel is characterized by the transmissivity $\eta \in (0,1)$, the noise mean photon number $N_B$, and the input-output relations 
\begin{align*}
\hat{b}
	&= \sqrt{\eta} \, \hat{a} + \sqrt{1-\eta} \, \hat{c}, \\
\hat{e}
	&= -\sqrt{1-\eta} \, \hat{a} + \sqrt{\eta} \, \hat{c},
\end{align*}
where $\hat{a}$ is the annihilation operator of the input mode, $\hat{c}$ is the annihilation operator of an environmental mode in a thermal state with mean photon number $N_B$ and density matrix $\rho_{N_B} \triangleq \sum_{n=0}^\infty \frac{N_B^n}{(N_B+1)^{n+1}} \ket{n}\bra{n}$, $\hat{b}$ is the annihilation operator of the accessible output mode, and $\hat{e}$ is the annihilation operator of the mode lost to the environment. For notation convenience, we define the effective thermal mean photon number as $N_T \triangleq (1-\eta)N_B$ and refer to this channel as an $(\eta, N_T)$-bosonic channel.

\begin{figure}[!htbp]
	\centering
    \includegraphics[width=0.7\linewidth]{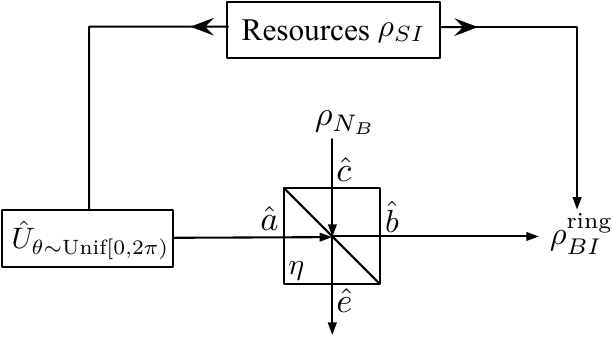}
    \caption{Quantum ring state.}
    \label{fig:ring_state}
\end{figure}

Consider now a bipartite Gaussian quantum state with density matrix $\rho_{SI}$, consisting of a signal arm $(S)$ and an idler arm $(I)$. Let $\rho_{BI}^\theta$ be the state obtained by applying a phase shift $\theta$ to the signal arm of $\rho_{SI}$ and transmitting the signal arm through an $(\eta,N_T)$-bosonic channel. As shown in Fig.~\ref{fig:ring_state}, a quantum ring state is ${\rho}_{BI}^{\textnormal{ring}} \triangleq \frac{1}{2 \pi}\int_{0}^{2\pi} \rho_{BI}^\theta d\theta$, which is the mixed state obtained by phase modulating the signal arm uniformly over $[0,2\pi)$, while leaving the idler arm unchanged and transmitting the modulated signal arm through a lossy thermal bosonic channel.

When the bipartite Gaussian quantum state is a \ac{TMSV} state, i.e., $\rho_{SI} \triangleq \ket{\psi} \bra{\psi}_{SI}$ where $\ket{\psi}_{SI} \triangleq \sum_{k=0}^{\infty} \sqrt{\frac{N_S^k}{(N_S+1)^{k+1}}}\, \ket{k}_S \ket{k}_I$, we refer to the ring state as a TMSV ring state ${\rho}_{BI}^\textnormal{TMSV-ring}$. We know from~\cite[(B5)]{Shi2020PracticalRoute} that the TMSV ring state is non-Gaussian but diagonal in the Fock basis, with coefficients involving a hypergeometric function. We extend this result in Proposition~\ref{prop:TMSV_ring_converge} by characterizing how fast an $m$-ary \ac{PSK}-modulated \ac{TMSV} state converges to the TMSV ring state as $m$ increases. This convergence result will be useful later on.

\begin{proposition}
\label{prop:TMSV_ring_converge}
Let $\rho_{BI,m}^\textnormal{TMSV}\triangleq \frac{1}{m}\sum_{k=0}^{m-1}\rho_{BI}^{\theta = \frac{2 \pi k}{m}}$ when $\rho_{SI}$ is a \ac{TMSV} state. If $N_T>\max\left\{\eta,\eta N_S-1\right\}$, then
\begin{multline*}
	1-F(\rho_{BI,m}^\textnormal{TMSV},{\rho}_{BI}^\textnormal{TMSV-ring})
	\\
\leq
	\frac{\frac{1}{2}\left(\frac{\eta N_S}{N_T+1}\right)^{m}}
	     {1-\left(\frac{\eta N_S}{N_T+1}\right)^{m}}
	+O\biggl(\left(\frac{\eta N_S}{N_T+1}\right)^{\frac{3m}{2}}\biggr),
\end{multline*}
where $F(\rho,\sigma) \triangleq \norm[1]{\sqrt{\rho} \sqrt{\sigma}}$ is the fidelity between quantum states $\rho$ and $\sigma$.
\end{proposition}
\begin{proof}
    See Appendix~\ref{sec:TMSV_ring_converge_proof}.
\end{proof}

Proposition~\ref{prop:TMSV_ring_converge} shows that the \ac{TMSV} ring state ${\rho}_{BI}^\textnormal{TMSV-ring}$ is closely approximated by the $m$-ary \ac{PSK}-modulated state $\rho_{BI,m}^\textnormal{TMSV}$. Specifically, $1-F(\rho_{BI,m}^\textnormal{TMSV},{\rho}_{BI}^\textnormal{TMSV-ring})$ converges exponentially fast to zero as $m$ increases. The convergence is particularly rapid in relevant regimes of low signal power or high thermal noise, for which $\frac{\eta N_S}{N_T+1}  \ll 1$.

When the bipartite Gaussian quantum state is a split coherent state, i.e., { $\rho_{SI}$} is obtained as the two outputs of a beamsplitter with transmissivity $\tau$, for which the inputs are a coherent state $\ket{\alpha}\bra{\alpha}$ and the vacuum state, we refer to the ring state as a coherent ring state ${\rho}_{BI}^\textnormal{coherent-ring}$. Analogous to Proposition~\ref{prop:TMSV_ring_converge}, we show in Proposition~\ref{prop:coherent_ring_converge} how fast an $m$-ary \ac{PSK}-modulated split coherent state converges to the coherent ring state as $m$ increases.
 
\begin{proposition}
\label{prop:coherent_ring_converge}
Let $\rho_{BI,m}^\textnormal{coherent}\triangleq \frac{1}{m}\sum_{k=0}^{m-1}\rho_{BI}^{\theta = \frac{2 \pi k}{m}}$ when $\rho_{SI}$ is a split coherent state.
If $N_T>\frac{\eta\tau\abs{\alpha}^2}{2}-1$, then
\begin{multline*}
1-F(\rho_{BI,m}^\textnormal{coherent},\rho_{BI}^\textnormal{coherent-ring})
\\
\leq
	\frac{N_T+1}{2\left(N_T+1\right)-\eta \tau\abs{\alpha}^2} 
	\;
	e^{\Upsilon}
	\frac{\Upsilon^m}{m!}
	+O\left(  \left(\frac{\Upsilon^{m}}{m!}\right)^{\frac{3}{2}}\right),
\end{multline*}
where $\Upsilon \triangleq \frac{2 \eta \tau\abs{\alpha}^2\left(N_T+1\right)^2}
			     {\left(2\left(N_T+1\right)-\eta \tau \abs{\alpha}^2\right) N_T^2}$.
\end{proposition}
\begin{proof}
See Appendix~\ref{sec:coherent_ring_converge_proof}.
\end{proof}

 Note that $1-F(\rho_{BI,m}^\textnormal{coherent},{\rho}_{BI}^\textnormal{coherent-ring})$ converges even faster to zero as $m$ increases than in Proposition~\ref{prop:TMSV_ring_converge}.

\section{PSK-Modulated Quantum Communication Over Lossy Thermal Bosonic Channel}
\label{sec:comm-comm}

\begin{figure}[!htbp]
	\centering
    \includegraphics[width=0.472\textwidth]{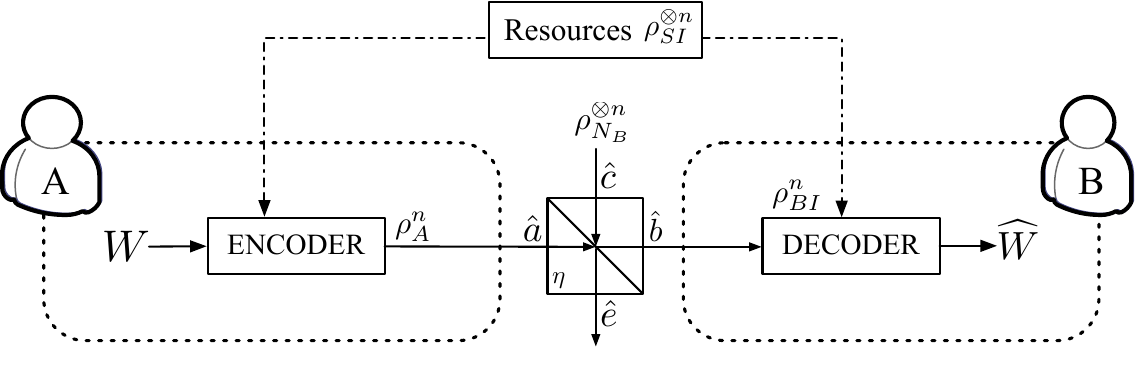}
    \caption{Communication model over a lossy thermal bosonic channel with pre-shared resources.}
    \label{fig:comm-model}
\end{figure}

Consider the communication model shown in Fig.~\ref{fig:comm-model}, in which a sender, Alice (A), transmits a classical message over an $(\eta,N_T)$-bosonic channel to a receiver, Bob (B), with the assistance of a pre-shared Gaussian resource. The resource also acts as a shared phase reference. The ultimate communication limit with unlimited shared entanglement is $C_\textnormal{EA} = g(N_S) + g(\eta N_S + N_T) - g\left( \frac{\mu_+ - 1}{2} \right) - g\left( \frac{\mu_- - 1}{2} \right)$, where $N_S$ denotes the mean photon number of the input state, $\mu_\pm \triangleq \sqrt{\left(N_S+\eta N_S+N_T+1\right)^2-4 \eta N_S\left(N_S+1\right)} \pm \left(N_T-(1-\eta) N_S\right)$, and $g(x) \triangleq (x+1) \log (x+1) - x \log x$~\cite{Bennett2002EntanglementassistedCapacity,Holevo2001EvaluatingCapacities}. The ultimate communication limit without entanglement is $C_\chi = g(\eta N_S + N_T) - g(N_T)$~\cite{Holevo2001EvaluatingCapacities,Giovannetti2014UltimateClassical}.

We focus on $m$-ary \ac{PSK} modulation, a structured and widely deployed modulation scheme in optical communication, which is commonly used in experiments~\cite{Shi2020PracticalRoute,Cox2023TransceiverDesigns,Fessatidis2026PracticalTwoWay,Hao2021EntanglementassistedCommunication}. Therein, Alice encodes messages $W$ into length-$n$ sequences of phases $\theta^n$ drawn from the finite set {$\left\{ \frac{2\pi k}{m} \right\}_{k=0}^{m-1}$} and transmits the resulting $n$-mode phase-modulated state through the channel. Bob estimates the message {$\widehat{W}$} by jointly measuring the channel output and the idler arm of the pre-shared resource, with the corresponding joint state denoted by $\bigotimes_{i=1}^n\rho_{BI}^{\theta_i}$.

The capacity of the lossy thermal bosonic channel when using $m$-ary \ac{PSK} is given by the Holevo information~\cite{Holevo2019QuantumSystems}
\begin{multline}
\label{eq:Holevo_info_m}
\chi_{m-\textnormal{PSK}}
\triangleq
	\vonH{\frac{1}{m} \sum_{k=0}^{m-1} \rho_{BI}^{\theta=\frac{2\pi k}{m}}}\\
-
	\frac{1}{m}\sum_{k=0}^{m-1}\vonH{\rho_{BI}^{\theta=\frac{2\pi k}{m}}},
\end{multline}
where $\vonH{\rho} \triangleq -\tr{\rho \log \rho}$ denotes the von Neumann entropy of a quantum state $\rho$ \footnote{ $\log$ denotes the natural logarithm throughout the present work.}. The second term on the right-hand side of~\eqref{eq:Holevo_info_m} can be evaluated directly when the pre-shared resource is a Gaussian state, since the von Neumann entropy is invariant under unitary operations and each individual phase-modulated state remains Gaussian~\cite{Serafini2023QuantumContinuous}. Analyzing the first term is much more challenging because it involves a non-Gaussian mixture, for which no closed-form eigendecomposition exists.

We shall derive new and remarkably simple closed-form lower bounds for~\eqref{eq:Holevo_info_m} as functions of $N_S$, $N_T$, and $\eta$, applicable to communication with both quantum and classical resources.
Our approach proceeds through two main steps:
\begin{inparaenum}[1)]
  \item we use perturbation theory~\cite{Grace2022PerturbationTheory} to bound the gap between $\chi_{m-\textnormal{PSK}}$  and the Holevo information of the ring state
\begin{align*}
	\chi_{\textnormal{ring}}
\triangleq
	\vonH{\rho_{BI}^\textnormal{ring}}-\frac{1}{2\pi}\int_{0}^{2\pi}\vonH{\rho_{BI}^{\theta}} d \theta
\end{align*} as a function of $m$;
   \item we develop tight lower bounds for $\chi_{\textnormal{ring}}$.
\end{inparaenum}

The Holevo information $\chi_{\textnormal{ring}}$ also captures the limits of quantum reading of optical memory cells~\cite{Tanggara2024QuantumoptimalInformation}, in which a classical message is encoded in the attenuation or phase shifts of cells probed by a quantum state.

\subsection{TMSV-Assisted Communication}
We first specialize the pre-shared resource to~\ac{TMSV} states. Because of the non-classical correlations between the signal and idler modes, such states are known to enhance achievable communication rates, particularly in the low-power regime~\cite{Cox2023TransceiverDesigns}.

\begin{proposition}
\label{prop:TMSV_m_penalty}
Consider a \ac{TMSV} pair $\rho_{SI}$.
If $N_T>\max\left\{\eta,\eta N_S-1\right\}$, then
\begin{align*}
	\chi_{m-\textnormal{PSK}}^\textnormal{TMSV}
\geq
	\chi_{\textnormal{ring}}^\textnormal{TMSV}
	 - \zeta(m),
\end{align*}
where the penalty term $\zeta(m)$ is a simple but slightly long algebraic expression explicitly given in~\eqref{eq:zeta_m} of Appendix~\ref{sec:TMSV_m_penalty_proof}.
\end{proposition}

\begin{proof}
See Appendix~\ref{sec:TMSV_m_penalty_proof}.
\end{proof}

\begin{theorem}
\label{thm:TMSV_ring_Holevo_info}
Consider a \ac{TMSV} pair $\rho_{SI}$.
If $N_T>\max\left\{\eta,\eta N_S-1\right\}$, then
{
\begin{multline}
\chi_{\textnormal{ring}}^\textnormal{TMSV}
\geq
	C_\textnormal{EA}\left( \eta,N_S,N_T \right)\\
	+g\left(N_T\right)
	-g\left(\eta N_S +N_T \right) \\
	+\left(2 \eta N_S+N_T\right)N_S \log \left(\frac{\left(N_T+1\right) \left(N_T-\eta \right)}{N_T \left(N_T-\eta+1\right)}\right) \\
	+N_S \log \left(\frac{N_T+1}{N_T-\eta+1}\right)\\
	+\eta  N_S \log \left(\frac{\left(N_T+1\right)^2 \left(N_T-\eta \right)}{N_T^2 \left(N_T-\eta+1\right)}\right).
	\label{eq:chi_TMSV_ring}
\end{multline}
}
\end{theorem}
\begin{proof}[Sketch of Proof]
We lower bound the entropy of the \ac{TMSV} ring state $\vonH{{\rho}_{BI}^\textnormal{TMSV-ring}} = -\sum_{n_1=0}^\infty \sum_{n_2=0}^\infty p\left( n_1,n_2 \right)\log p\left( n_1,n_2 \right)$, where $p\left( n_1,n_2 \right)$ denotes the joint probability mass function of the photon numbers in the two modes.
The distribution $p\left( n_1,n_2 \right)$ contains the hypergeometric function {${}_2F_1[a,b;c;z] \triangleq \sum_{k=0}^{\infty} \frac{(a)_k(b)_k}{(c)_k k!} z^k$}, where { $(x)_k = \frac{\Gamma(x+k)}{\Gamma(x)}$} denotes the Pochhammer symbol. We carefully lower bound the contribution { $-\sum_{n_1=0}^\infty \sum_{n_2=0}^\infty p\left( n_1,n_2 \right)\log {}_2F_1[n_1+1,n_2+1;1;z]$} as detailed in Appendix~\ref{sec:TMSV_ring_Holevo_info_proof}.
\end{proof}

In contrast to prior results in the literature, e.g.,~\cite[(B7)]{Shi2020PracticalRoute} and~\cite[Lemma A1]{DiCandia2021TwowayCovert}, our approach does not rely on asymptotic assumptions or Taylor expansions in the regime of low transmitted mean photon number $N_S$, high effective thermal mean photon number $N_T$, or low transmissivity $\eta$, which can be ambiguous in the presence of cross terms such as $N_T \cdot N_S$. The lower bound in~\eqref{eq:chi_TMSV_ring} provides an achievable communication rate in closed-form for the TMSV ring state that is valid across a broad parameter range.

\begin{figure}[!htbp]
    \centering
    \includegraphics[width=0.98\linewidth]{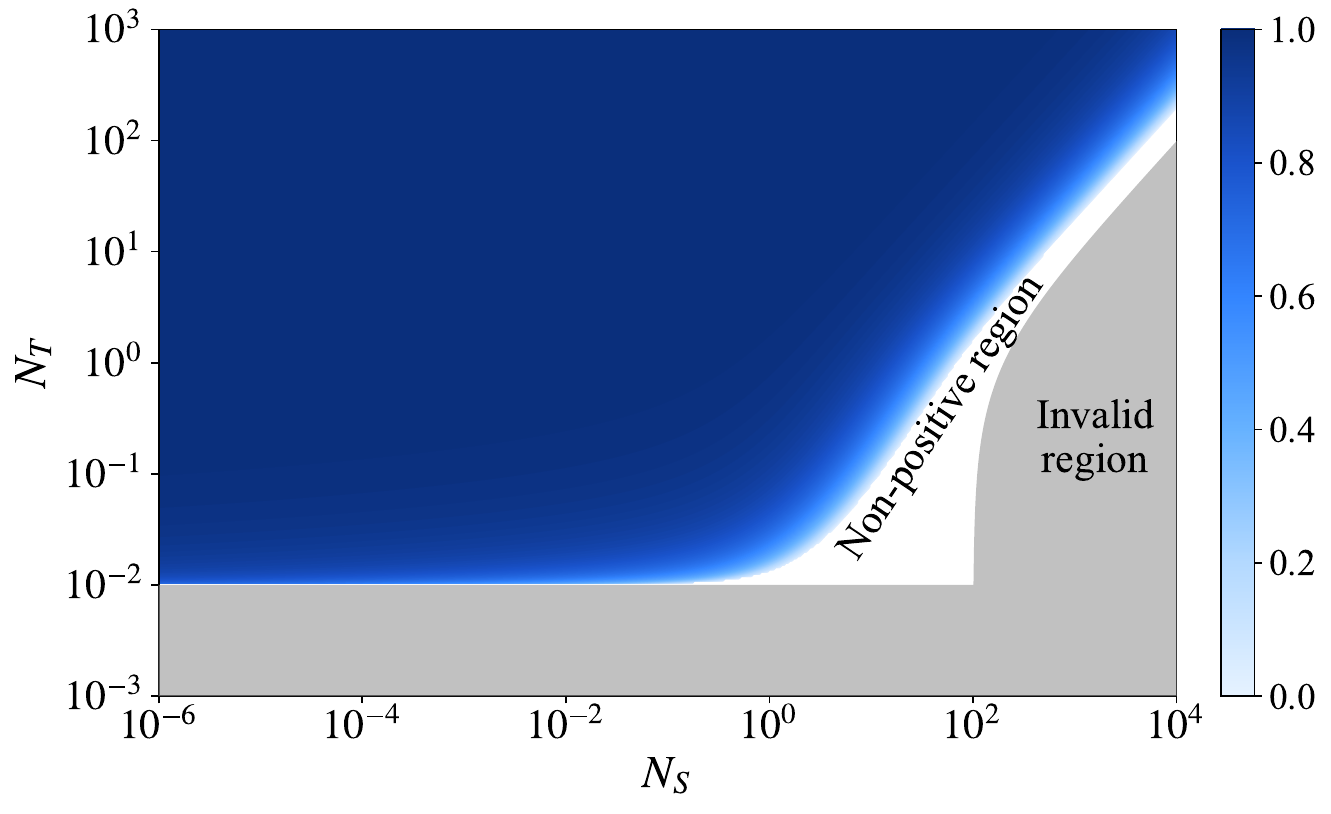}
    \caption{Ratio of the achievable rate in Theorem~\ref{thm:TMSV_ring_Holevo_info} to the entanglement-assisted capacity $C_\textnormal{EA}$ for channel transmissivity $\eta = 0.01$ (higher is better).}
    \label{fig:p2_rate_over_Cea}
\end{figure}

Fig.~\ref{fig:p2_rate_over_Cea} illustrates the usefulness of Theorem~\ref{thm:TMSV_ring_Holevo_info} by showing the ratio of our lower bound in~\eqref{eq:chi_TMSV_ring} to the entanglement-assisted capacity $C_\textnormal{EA}$. The grey invalid region at the bottom-right corner captures the parameter regime in which the convergence result of Proposition~\ref{prop:TMSV_ring_converge} does not apply. The solid white region captures the parameter regime in which our lower bound in~\eqref{eq:chi_TMSV_ring} is non-positive and therefore not useful. The large shaded region on the left captures the parameter regime in which~\eqref{eq:chi_TMSV_ring} is useful. In particular, $\chi_{\textnormal{ring}}$ approaches $C_\textnormal{EA}$ across a broad range of $N_S$ and $N_T$ as $N_S$ decreases.

\begin{figure}[!htbp]
    \centering
    \includegraphics[width=0.98\linewidth]{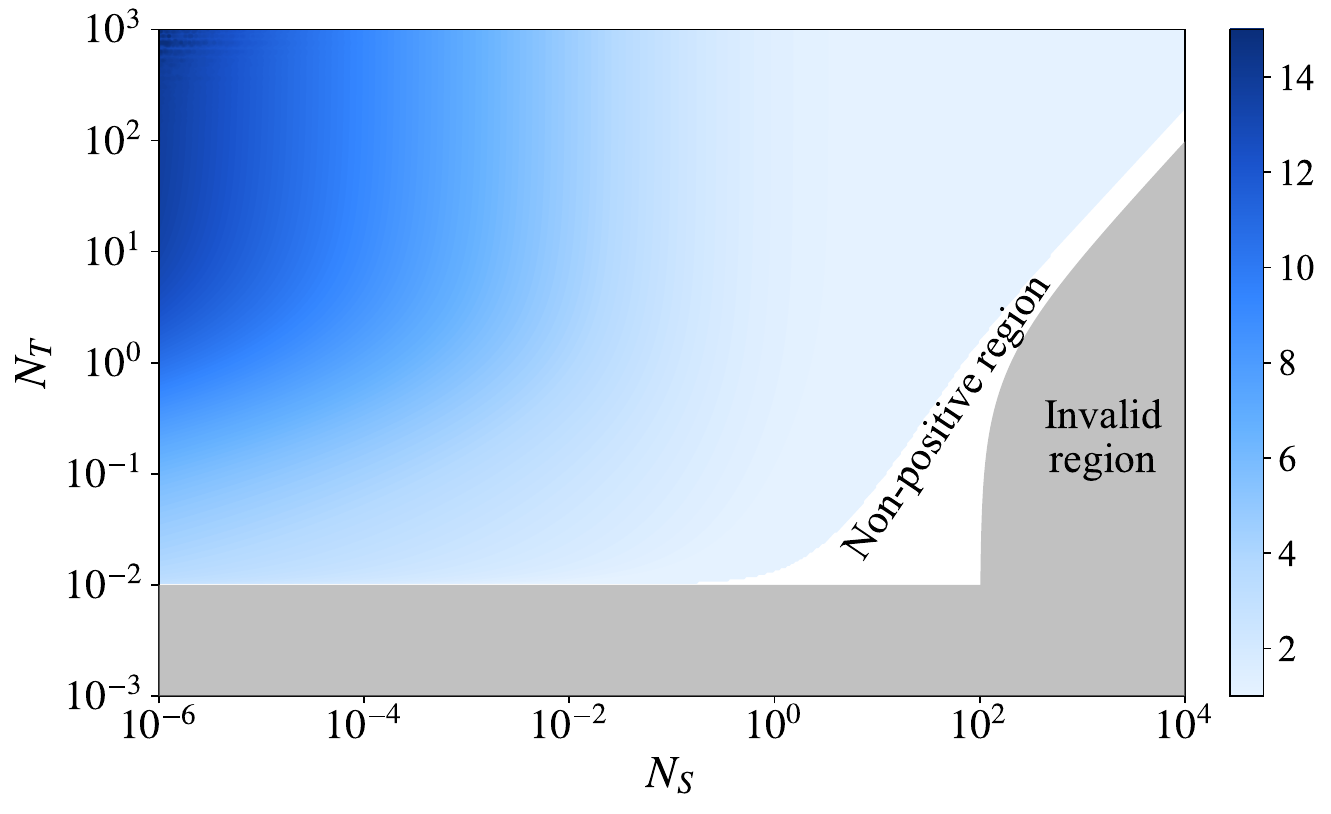}
    \caption{Ratio of the achievable rate in Theorem~\ref{thm:TMSV_ring_Holevo_info} to the classical capacity $C_\chi$ for channel transmissivity $\eta = 0.01$ (higher is better).}
    \label{fig:p2_rate_over_Cchi}
\end{figure}

Fig.~\ref{fig:p2_rate_over_Cchi} illustrates the benefit of entanglement by showing the ratio of our lower bound in~\eqref{eq:chi_TMSV_ring} to the classical capacity $C_\chi$. As expected, the strongest quantum advantage occurs when $N_S$ is small and $N_T$ is large, consistent with previously characterized asymptotic behavior reported in~\cite{Cox2023TransceiverDesigns}.

\begin{figure}[!htbp]
    \centering
    \includegraphics[width=0.98\linewidth]{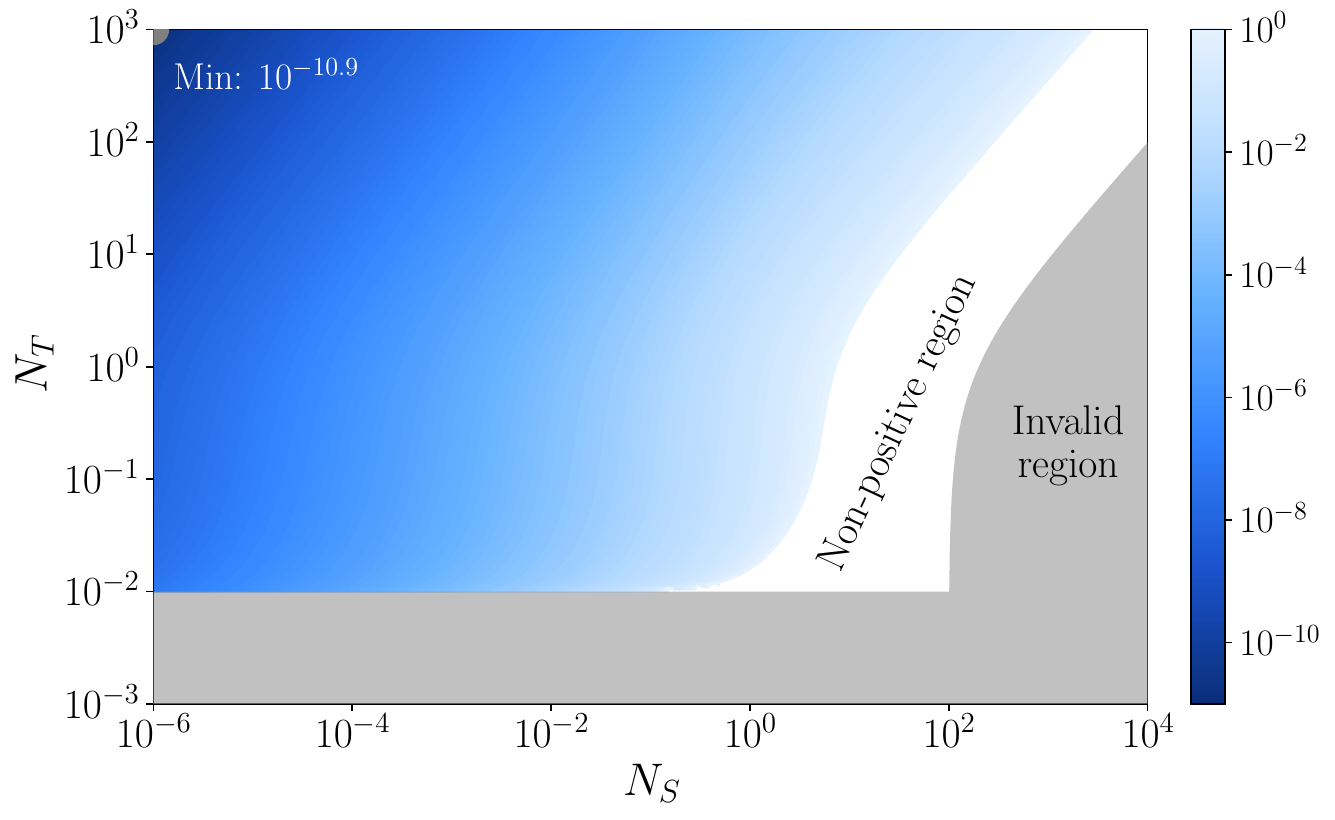}
    \caption{BPSK modulation rate penalty (lower is better), expressed as $\frac{\zeta(m=2)}{\chi_\textnormal{ring}^\textnormal{TMSV}-\zeta(m=2)}$ for channel transmissivity $\eta = 0.01$.}
    \label{fig:TMSV_BPSK_penalty_0d10}
\end{figure}

Finally, Fig.~\ref{fig:TMSV_BPSK_penalty_0d10} illustrates the usefulness of Proposition~\ref{prop:TMSV_m_penalty} by plotting the relative rate penalty $\frac{\zeta(m=2)}{\chi_\textnormal{ring}^\textnormal{TMSV}-\zeta(m=2)}$ incurred by~\ac{BPSK} modulation. The relative rate penalty remains negligible over a broad region of interest, thereby illustrating the benefit of our approach in capturing the performance not only of the ring state but also of finite $m$-ary \ac{PSK} modulation schemes.


The regime of low transmitted mean photon number $N_S \ll 1$ is where quantum advantages are substantial. In this regime, taking the limit $N_S \to 0^+$ offers additional insights into the relative impacts of the channel parameters on communication rates. Taking the limit in Theorem~\ref{thm:TMSV_ring_Holevo_info}, we obtain the following result.

\begin{proposition}
\label{prop:TMSV_low_power_one_way}
If $N_T > \eta$ and in the limit of ${N_S \to 0^+}$
\begin{multline*}
\chi_{\textnormal{ring}}^\textnormal{TMSV}
\geq
	-\frac{\eta N_S \log N_S}{N_T+1}\\
	+N_S
		\biggl(    \left(N_T+\eta \right)\log \frac{(N_T+1)(N_T-\eta)}{N_T(N_T-\eta+1)}
		\\
		+\frac{\eta}{N_T+1} \left( 1+ \log \frac{(N_T+1)^2}{N_T(N_T-\eta+1)}\right)
		\biggr) \\
	+O\left(-N_S^2 \log N_S\right).
\end{multline*}
\end{proposition}

The leading order in Proposition~\ref{prop:TMSV_low_power_one_way} matches the leading order of the entanglement-assisted capacity, $C_\textnormal{EA}$, thereby confirming that \ac{PSK} modulation is optimal in the regime $N_S \ll 1$~\cite{Shi2020PracticalRoute} but with minimal assumptions on $N_T$ and $\eta$. Although the subsequent terms of the expansion are not optimal, they reveal the complex interplay between the channel parameters. A byproduct of our analysis is that the number of consumed \ac{TMSV} states scales linearly with the number of channel uses. More precisely, the scheme consumes $g(N_S)$ ebits per channel use.

\begin{figure}[!htbp]
    \centering
    \includegraphics[width=0.98\linewidth]{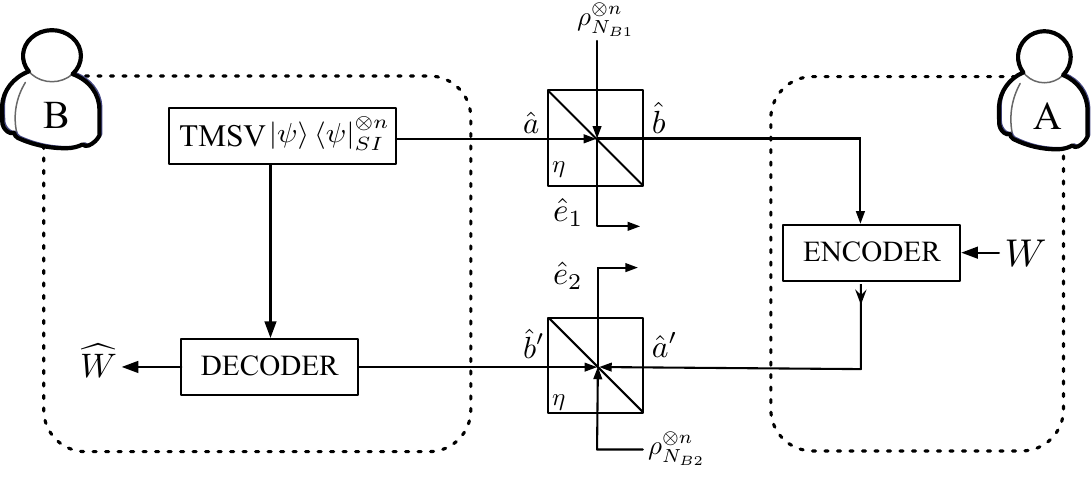}
    \caption{Round-trip communication model.}
    \label{fig:round_trip_comm}
\end{figure}

Our approach readily extends to the round-trip communication model illustrated in Fig.~\ref{fig:round_trip_comm}~\cite{DiCandia2021TwowayCovert,Fessatidis2026PracticalTwoWay}, in which the signal arm of the \ac{TMSV} state is transmitted from Bob to Alice through a forward $(\eta,(1-\eta)N_{B1})$-bosonic channel, phase-modulated by Alice, and then returned to Bob through a backward $(\eta,(1-\eta)N_{B2})$-bosonic channel. This model better captures experimental realizations of \ac{TMSV}-assisted communication~\cite{Hao2021EntanglementassistedCommunication,Fessatidis2026PracticalTwoWay}.

\begin{proposition}
\label{prop:TMSV_ring_Holevo_info_roundtrip}
Let $\eta^\prime \triangleq \eta^2$, and $N_T^\prime \triangleq (1-\eta)\eta N_{B1}+(1-\eta)N_{B2}$. If $N_T^\prime>\max\left\{\eta^\prime,\eta^\prime N_S-1\right\}$, then
\begin{multline*}
\chi_{\textnormal{ring,round-trip}}^\textnormal{TMSV}
\geq
	C_\textnormal{EA} \left(\eta^\prime,N_S,N_T^\prime \right)\\
	+g\left(N_T^\prime\right)
	-g\left(\eta^\prime N_S +N_T^\prime \right) \\
	+\left(2 \eta^\prime N_S+N_T^\prime\right)N_S \log \left(\frac{\left(N_T^\prime+1\right) \left(N_T^\prime-\eta^\prime \right)}{N_T^\prime \left(N_T^\prime-\eta^\prime+1\right)}\right) \\
	+N_S \log \left(\frac{N_T^\prime+1}{N_T^\prime-\eta^\prime+1}\right) \\
	+\eta^\prime  N_S \log \left(\frac{\left(N_T^\prime+1\right)^2 \left(N_T^\prime-\eta^\prime \right)}{(N_T^\prime)^2 \left(N_T^\prime-\eta^\prime+1\right)}\right).
\end{multline*}
\end{proposition}
\begin{proof}[Sketch of Proof]
The covariance matrix of Bob's Gaussian state in the round-trip model takes the same form as in the one-way model with $\eta$ and $N_T$ replaced by $\eta^\prime$ and $N_T^\prime$, respectively. Consequently, the analysis for the one-way communication applies directly. The complete derivation is provided in Appendix~\ref{sec:round_trip_comm_proof}.
\end{proof}

Taking again the limit $N_S \to 0^+$, we obtain the following result.

\begin{proposition}
\label{prop:TMSV_low_power_roundtrip}
If $N_T^\prime>\eta^\prime$ and in the limit of ${N_S \to 0^+}$
\begin{multline*}
	\chi_{\textnormal{ring,round-trip}}^\textnormal{TMSV}
\geq
	-\frac{\eta^2 N_S \log N_S}{(1-\eta)\eta N_{B1}+(1-\eta)N_{B2}+1}\\
	+O(N_S).
\end{multline*}
\end{proposition}

The degradation of the achievable rate compared to the one-way model arises from the additional transmissivity factor and the accumulation of thermal noise from both forward and backward channels.

\subsection{Classical Communication}
\begin{figure}[!htbp]
    \centering
    \includegraphics[width=0.99\linewidth]{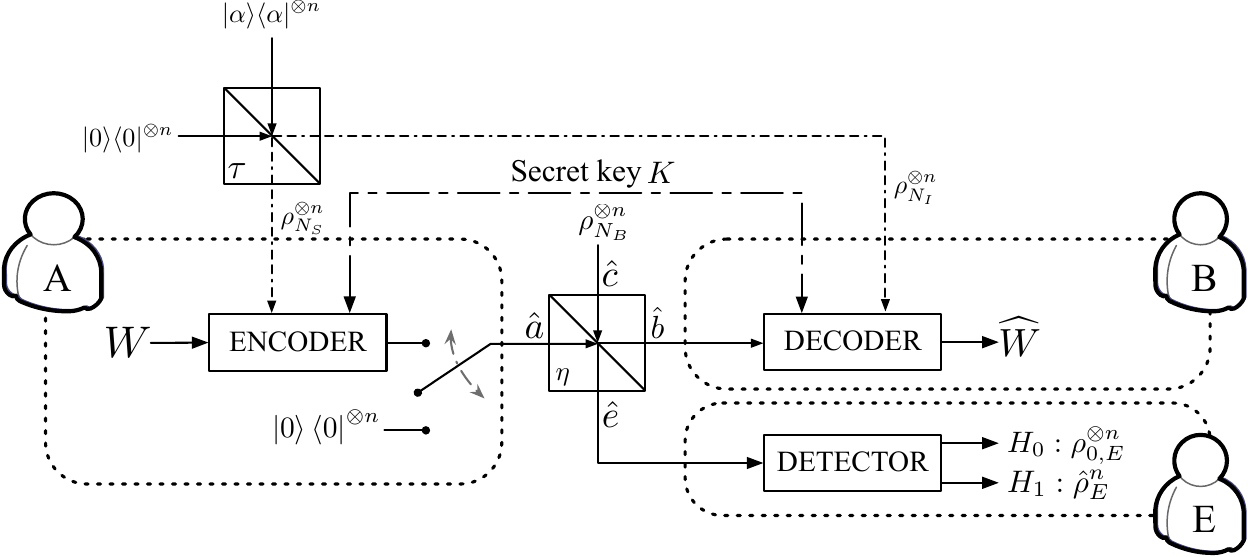}
    \caption{Communication model over a lossy thermal bosonic channel using coherent states as resources.}
    \label{fig:comm_model_coherent}
\end{figure}

We now specialize the pre-shared resource to coherent states, which model ideal lasers used in conventional optical communication. As shown in Fig.~\ref{fig:comm_model_coherent}, the bipartite resource is created by sending a coherent state $\ket{\alpha}\bra{\alpha}$ onto a beamsplitter with transmissivity $\tau$, while keeping the other input in the vacuum state. In the setup, the mean photon number of the signal arm is $N_S=\tau \abs{\alpha}^2$. As seen in the proof of Proposition~\ref{prop:coherent_ring_converge}, the received joint state $\rho_{BI}$ is a product state in system $B$ and system $I$, with the reduced state $\rho_I$ pure, which makes the analysis simpler than for \ac{TMSV}.

\begin{proposition}
\label{prop:coherent_m_penalty}

Consider a split coherent state $\rho_{SI}$.
If $N_T>\frac{\eta\tau\abs{\alpha}^2}{2}-1$, then
\begin{align*}
	\chi_{m-\textnormal{PSK}}^\textnormal{coherent}
\geq
	\chi_{\textnormal{ring}}^\textnormal{coherent}
	 - \xi(m),
\end{align*}
where the penalty term $\xi(m)$ is explicitly given in~\eqref{eq:xi_m_coherent} of Appendix~\ref{sec:coherent_m_penalty_proof}.
\end{proposition}

\begin{proof}[Sketch of Proof]
Similar to Proposition~\ref{prop:TMSV_m_penalty}, we apply perturbation theory for the von Neumann entropy~\cite{Grace2022PerturbationTheory}. Detailed derivations are provided in Appendix~\ref{sec:coherent_m_penalty_proof}.
\end{proof}

\begin{theorem}
\label{thm:coherent_ring_Holevo_info}
Consider a split coherent state $\rho_{SI}$.
If $N_T>\frac{\eta\tau\abs{\alpha}^2}{2}-1$, then
\begin{multline}
\chi_{\textnormal{ring}}^\textnormal{coherent}
\geq
	\eta \tau \abs{\alpha}^2
	  \left( \frac{1}{N_T+1} + \log \frac{N_T+1}{N_T} \right)\\
	 -\left( N_T+\eta\tau\abs{\alpha}^2 \right)\log\left( 1+ \frac{\eta \tau \abs{\alpha}^2}{N_T\left(N_T+1\right)}\right).
	 \label{eq:chi_coherent}
\end{multline}
\end{theorem}
\begin{proof}
  See Appendix~\ref{sec:coherent_ring_Holevo_info_proof}.
\end{proof}

While prior work~\cite{Gagatsos2020CovertCapacity, Wang2022CharacterizationCovert} only derived achievable communication rates with \ac{QPSK} modulation in the asymptotic regime of vanishingly small $N_S$, our bound~\eqref{eq:chi_coherent} provides an achievable communication rate in closed-form for the coherent ring state that holds over a large range of channel parameters without relying on asymptotic assumptions. Our bound also generalizes~\cite[Appendix G]{Tanggara2024QuantumoptimalInformation}, which also relies on a Taylor expansion.

\begin{figure}[!htbp]
    \centering
    \includegraphics[width=0.98\linewidth]{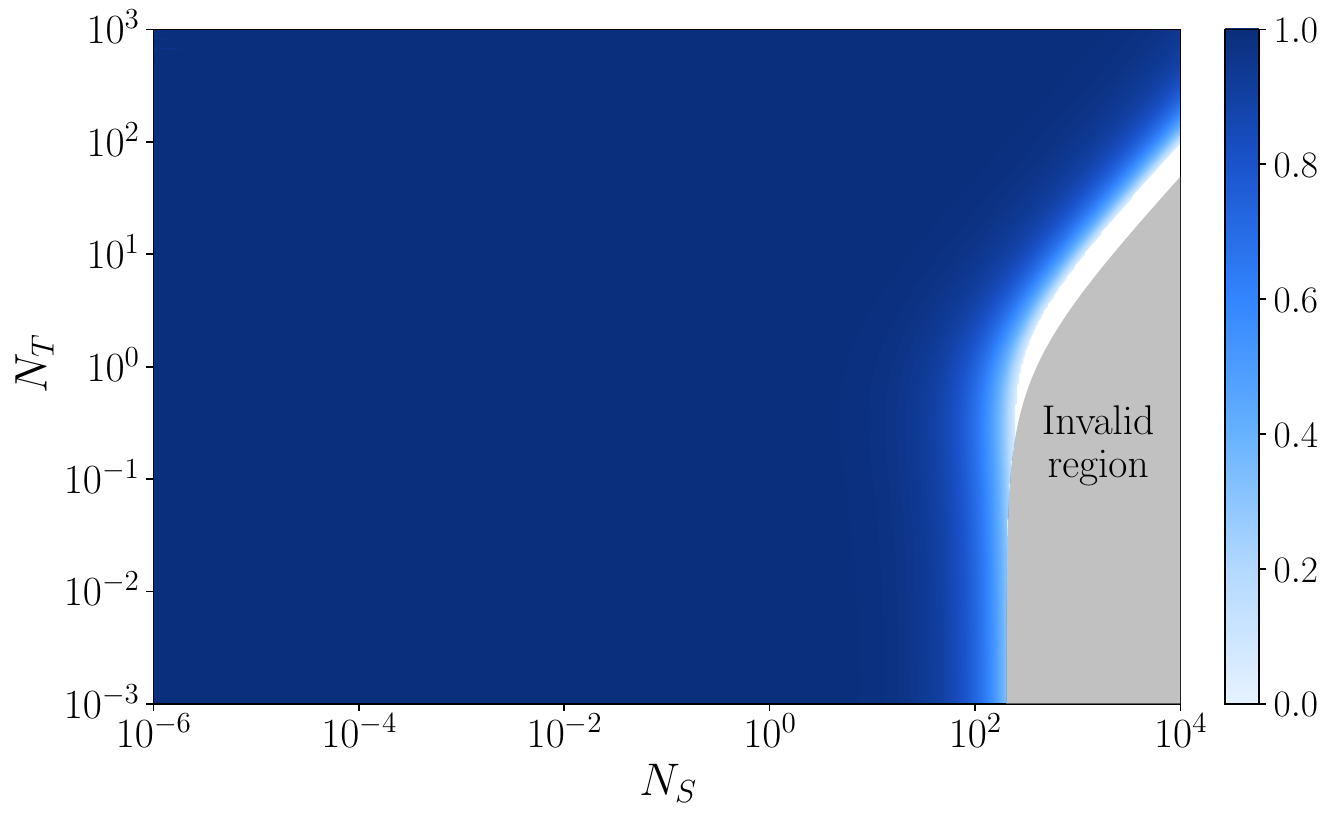}
    \caption{Ratio of the achievable rate in Theorem~\ref{thm:coherent_ring_Holevo_info} to the classical capacity $C_\chi$ for channel transmissivity $\eta = 0.01$ (higher is better).}
    \label{fig:coherent_rate_over_C_chi}
\end{figure}

Fig.~\ref{fig:coherent_rate_over_C_chi} illustrates the usefulness of Theorem~\ref{thm:coherent_ring_Holevo_info} by showing the ratio of our lower bound in~\eqref{eq:chi_coherent} to the classical capacity $C_\chi$. The grey invalid region in the bottom-right corner captures the parameter regime in which the convergence result of Proposition~\ref{prop:coherent_ring_converge} does not apply. The shaded region on the left captures the parameter regime in which $\chi_{\textnormal{ring}}$ approaches $C_\chi$ and is even larger than the corresponding region for the~\ac{TMSV} ring state in Fig.~\ref{fig:p2_rate_over_Cea}.

\begin{figure}[!htbp]
    \centering
    \includegraphics[width=0.98\linewidth]{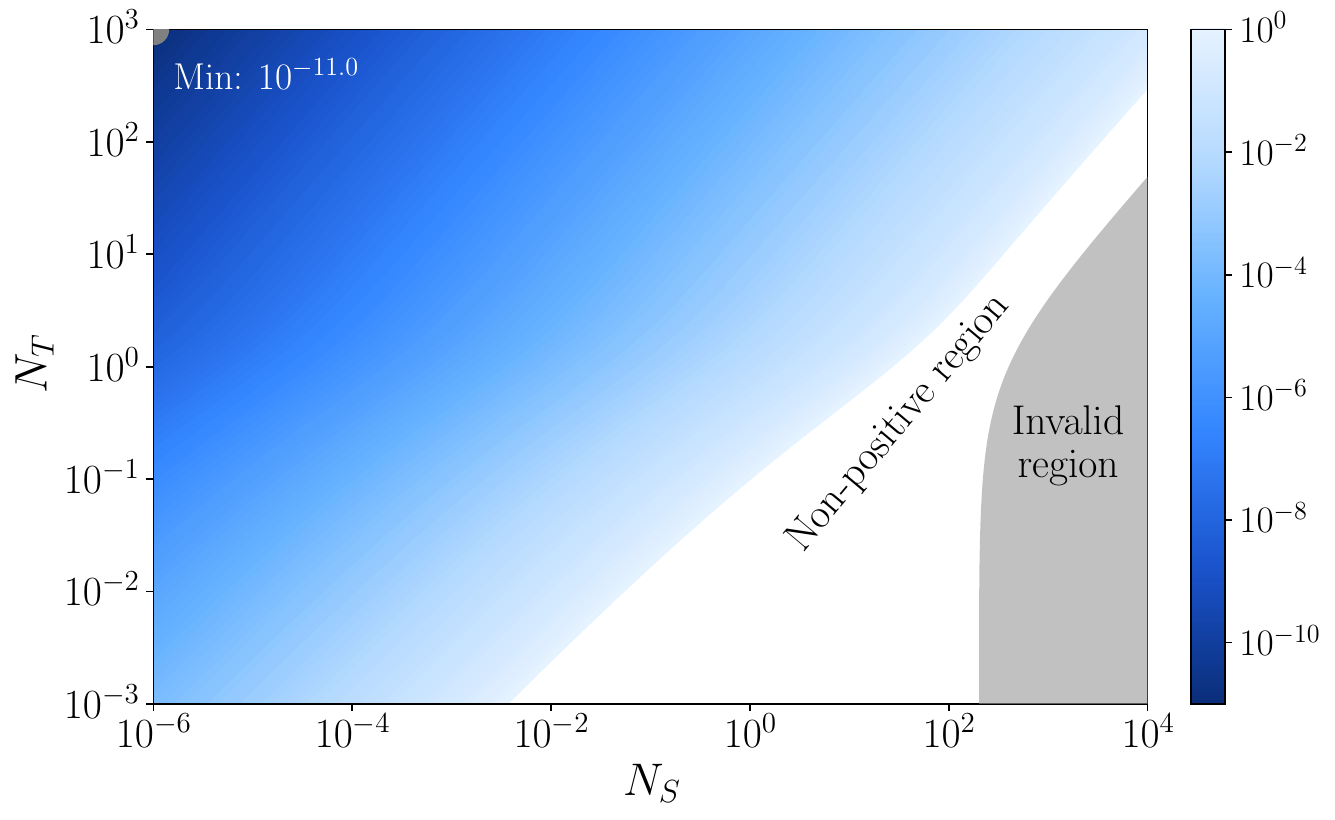}
    \caption{BPSK modulation rate penalty (lower is better), expressed as $\frac{\xi(m=2)}{{\chi_\textnormal{ring}^\textnormal{coherent}-\xi(m=2)}}$ for channel transmissivity $\eta = 0.01$.}
    \label{fig:coherent_BPSK_penalty_0d10}
\end{figure}

Finally, Fig.~\ref{fig:coherent_BPSK_penalty_0d10} illustrates the usefulness of Proposition~\ref{prop:coherent_m_penalty} by plotting the relative rate penalty $\frac{\xi(m=2)}{\chi_\textnormal{ring}^\textnormal{coherent}-\xi(m=2)}$ incurred by~\ac{BPSK} modulation. The relative rate penalty remains negligible over a broad region of interest, thereby illustrating again the benefit of our approach in capturing again not only the performance of the ring state but also finite $m$-ary \ac{PSK} modulation schemes.

Taking the limit $N_S = \tau \abs{\alpha}^2 \to 0^+$ in Theorem~\ref{thm:coherent_ring_Holevo_info}, we obtain the following asymptotic approximation.

\begin{proposition}
\label{prop:coherent_low_power_one_way}
In the limit of ${N_S \to 0^+}$
\begin{multline*}
\chi_{\textnormal{ring}}^\textnormal{coherent}
\geq
	 \eta N_S \log \left(1+\frac{1}{N_T}\right) \\
	-\frac{2 N_T+1}{2 N_T \left(N_T+1\right)^2} \eta ^2 N_S^2
	+O\left(N_S^3\right).
\end{multline*}
\end{proposition}

The leading order of Proposition~\ref{prop:coherent_low_power_one_way} matches that of the classical capacity $C_\chi$. Compared to other schemes~\cite{Bullock2020FundamentalLimits,Gagatsos2020CovertCapacity} that achieve the same leading term, the ring-state analysis is perhaps more direct. The analysis for the round-trip communication model with coherent ring states is similar to that in Proposition~\ref{prop:TMSV_ring_Holevo_info_roundtrip} and is therefore omitted for brevity.

\section{Quantum Covert Communication}
\label{sec:application_covert}

We further apply quantum ring states to study one-way~\cite{Bash2015QuantumsecureCovert,Gagatsos2020CovertCapacity,Wang2022CharacterizationCovert,Cox2023TransceiverDesigns,Wang2024ResourceefficientEntanglementassisted} and round-trip~\cite{DiCandia2021TwowayCovert,Fessatidis2026PracticalTwoWay} quantum covert communications. Covert communication extends the communication setup studied in Sec.~\ref{sec:comm-comm} by requiring the communication to remain not only reliable for a legitimate receiver but also undetectable by an eavesdropper observing the environment mode. Covertness is typically achieved by operating in the regime of low transmitted mean photon number, so that the transmission remains statistically indistinguishable from background noise. This regime coincides with the operating conditions under which ring states exhibit optimal communication performance.

\subsection{One-Way Covert Communication}
\label{sec:one_way_covert}

\begin{figure}[!htbp]
    \centering
    \includegraphics[width=0.98\linewidth]{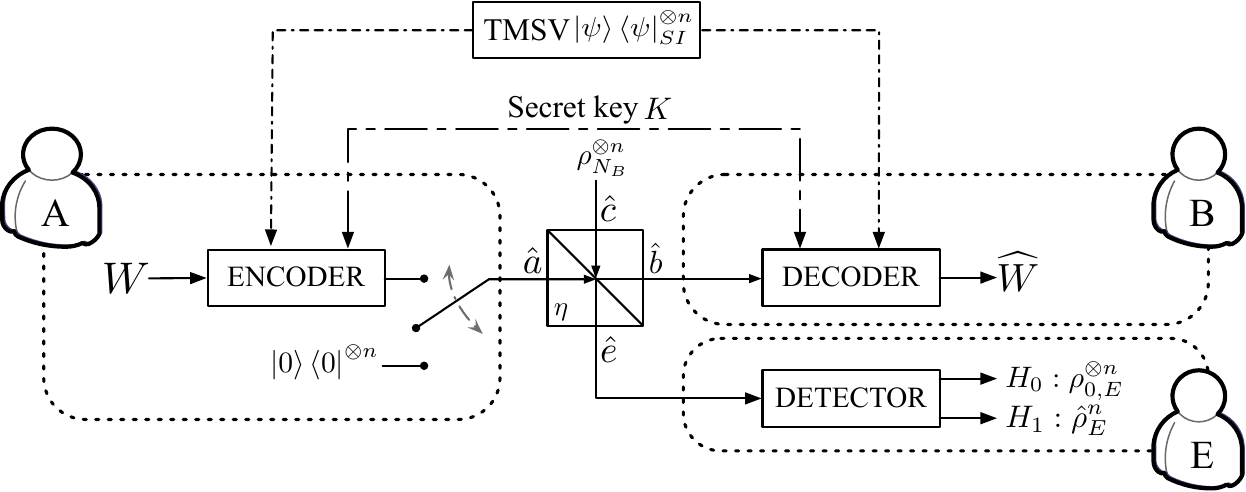}
    \caption{One-way covert communication model with TMSV ring states.}
    \label{fig:one_way_covert_TMSV}
\end{figure}

We first consider one-way covert communication using \ac{TMSV} ring states, as illustrated in Fig.~\ref{fig:one_way_covert_TMSV}. In this model, a transmitter (Alice) has access to the signal arm of \ac{TMSV} states and a secret key to transmit a message to a receiver (Bob) observing the output of the $(\eta,(1-\eta) N_B)$-bosonic channel and decoding with the assistance of the idler arm of the same \ac{TMSV} states and the secret key. The communication is eavesdropped on by an adversary (Eve) who should remain oblivious to the presence of the communication.

More specifically, Alice encodes a message $w \in [1;M]$ and a secret key $k \in [1;L]$ into a sequence of phase shifts $\theta^n$ that modulate the signal arm of the $n$ \ac{TMSV} states. This encoding defines a set of quantum channels $\left\{\mathcal{E}_{ {S}^n \rightarrow A^n}^{(w, k)}\right\}_{w,k}$ acting on the signal arm, which we assume are publicly known. Bob's decoder consists of a set of \acp{POVM} $\left\{\Pi^{(w, k)}_{B^n I^n} \right\}_{w,k}$, which act on the joint quantum state of the received signal in system $B^n$ and the idler arm of the \ac{TMSV} states $I^n$. Let  $\mathcal{E}_{A \rightarrow E}$ represents the $(1-\eta,\eta N_B)$-bosonic channel connecting Alice to Eve. In the presence of communication, Eve observes the state
\begin{align*}
	\hat{\rho}_{E}^n = \frac{1}{M L}\sum_{w=1}^M \sum_{k=1}^L \mathcal{E}_{A \rightarrow E}^{\otimes n} \; \rho_{A^n}^{(w,k)}	
\end{align*}
where $\rho^{(w, k)}_{A^n}= \operatorname{tr}_{I^n}\left(\left( \mathcal{E}_{ {S}^n \rightarrow A^n}^{(w, k)} \otimes \hat{I}_{I^n} \right) \rho_{SI}^{\otimes n} \right)$ and $\hat{I}$ denotes the identity operator. In contrast, in the absence of communication, Eve observes the state
\begin{align*}
	{\rho}_{0,E}^{\otimes n} = \mathcal{E}_{A \rightarrow E}^{\otimes n} \; \ket{0}\bra{0}^{\otimes n}.
\end{align*}

A code is $\varepsilon$-reliable if $\mathbb{E}_{{K}}[\mathbb{P}(\widehat{W} \neq W \mid {K})] \leq \varepsilon$. A code is $\delta$-covert if $\frac{1}{2}\norm[1]{\smash{\hat{\rho}_{E}^n - {\rho}_{0,E}^{\otimes n}}} \leq \delta$ holds. Indeed, for any binary hypothesis test attempting to distinguish the states $\hat{\rho}_{E}^n$ and $\rho_{0, E}^{\otimes n}$, the probability of false alarm $\alpha$ and the probability of missed detection $\beta$ satisfy $1 \geqslant \alpha+\beta \geqslant 1-\frac{1}{2}\left\|\smash{\hat{\rho}_{E}^n-\rho_{0, E}^{\otimes n}}\right\|_1$, where the lower bound can be achieved by the Helstrom test. A small $\delta$ therefore constrains the detection performance to be no better than a random guess. See Appendix~\ref{sec:covert_lemma} for the technical details of the design of $\varepsilon$-reliable and $\delta$-covert codes.

\begin{theorem}
\label{thm:TMSV_one_way_covert}
There exist $\varepsilon$-reliable and $\delta$-covert codes operating on $n$ phase-modulated \ac{TMSV} states without secret keys such that 
\begin{align*}
&	\lim_{n \rightarrow \infty} \frac{\log M}{\sqrt{n}\log n}
\geq
	\frac{\eta}{1-\eta}
	\frac{\sqrt{\frac{\eta}{1-\eta}N_T(\frac{\eta}{1-\eta}N_T+1)}}{N_T+1} \; \delta.&
\end{align*}
\end{theorem}
\begin{proof}[Sketch of Proof]
The covertness constraint can be upper bounded as
\begin{align*}
	\frac{1}{2} \norm[1]{\hat{\rho}_{E}^n- \rho_{0,{E}}^{\otimes n}} \leq \frac{1}{2} \norm[1]{\hat{\rho}_{E}^n-\rho_{n,E}^{\otimes n}}+\frac{1}{2}\norm[1]{{\rho}_{n,{E}}^{\otimes n}-\rho_{0,E}^{\otimes n}},
\end{align*}
where $\rho^{\otimes n}_{n,E} \triangleq \mathcal{E}_{A \rightarrow E}^{\otimes n} \; \rho_{A^n}$ is a thermal state. Since thermal states are invariant under phase shifts, the first trace distance $\frac{1}{2}\norm[1]{\smash{\hat{\rho}_{E}^n-\rho_{n,E}^{\otimes n}}}$ is zero. As shown in Appendix~\ref{sec:TMSV_one_way_covert_proof}, the second trace distance $\frac{1}{2}\norm[1]{\smash{{\rho}_{n,E}^{\otimes n}-\rho_{0,E}^{\otimes n}}}$ is less than $\delta$ if $N_S \leq \frac{1}{\sqrt{n}}\frac{\delta}{\sqrt{C_2}}$, where $C_2 = \frac{(1-\eta)^2}{4\frac{\eta}{1-\eta}N_T \left(\frac{\eta}{1-\eta}N_T +1\right)}$. The result of Theorem~\ref{thm:TMSV_one_way_covert} is obtained by evaluating the leading term of Proposition~\ref{prop:TMSV_low_power_one_way} at the maximum permissible signal power $N_S = \frac{1}{\sqrt{n}}\frac{\delta}{\sqrt{C_2}}$. The complete derivation is provided in Appendix~\ref{sec:TMSV_one_way_covert_proof}.
\end{proof}

The lower bound in Theorem~\ref{thm:TMSV_one_way_covert} is sometimes called the covert throughput; the lower bound is $\sqrt{2 \delta}$ times higher than the equivalent result in~\cite[(30)]{Gagatsos2020CovertCapacity}. This difference stems from the distinct covertness constraints employed: the present work uses the trace distance while~\cite{Gagatsos2020CovertCapacity} adopts a relative entropy-based criterion. Both results show that the number of reliable nats under a covertness constraint scales as $O(\sqrt{n}\log n)$. A subtle difference is that our approach based on ring states shows that this is achievable using only $n$ \ac{TMSV} states while~\cite{Gagatsos2020CovertCapacity} potentially uses a much larger number of entangled states. The result of Theorem~\ref{thm:TMSV_one_way_covert} also extends to situations in which the \ac{TMSV} states are not ideal. In particular, as shown next, we are able to consider a \ac{TMSV} source that imparts non-\ac{i.i.d.} phases across the $n$ channel uses, which can happen in practice because of the finite linewidth of lasers.

\begin{proposition}
\label{prop:one_way_cover_imperfect_TMSV}
The result of Theorem~\ref{thm:TMSV_one_way_covert} holds in the presence of non-ideal \ac{TMSV} states whose phases are correlated across channel uses.
\end{proposition}

\begin{proof}[Sketch of Proof]
Reliability is unaffected by phase imperfections because Bob retains the idler arm of the \ac{TMSV} states, which remains entangled with the signal arm modulated by Alice. Regardless of whether the source phases across $n$ channel uses are \ac{i.i.d.} or correlated, Bob can perform joint measurements on the idler and the received signal to decode Alice's message.

We next verify that covertness is preserved in the presence of phase drift. Let the imperfect \ac{TMSV} phase be $\phi + \tilde{\phi}$, where $\phi \sim U[0,2\pi)$ and $\tilde{\phi}$ denotes a phase drift that may depend on $\phi$. We assume a more powerful Eve who has access to the phase reference in system $E^\prime$ and receives the state
$\rho_{E{E^\prime}}^\textnormal{drift}=
	\frac{1}{m}
	\sum_{k=0}^{m-1}
	 \left(\hat{U}_{\tilde{\phi}}\otimes\hat{I}_{E^\prime} \right)
	\rho_{E{E^\prime}}^{\theta = \frac{2\pi k}{m}}
	 \left(\hat{U}_{\tilde{\phi}}^\dagger\otimes\hat{I}_{E^\prime} \right)$.
Because the trace distance is invariant under unitary operations and because the phase-modulated state $\frac{1}{m}
	\sum_{k=0}^{m-1}
	\rho_{E{E^\prime}}^{\theta = \frac{2\pi k}{m}}$ converges in trace distance to Eve's state $\rho_{E E^\prime}^\textnormal{ideal}$ corresponding to an ideal \ac{TMSV} source (see Proposition~\ref{prop:TMSV_ring_converge}), we obtain
	$\frac{1}{2}\norm[1]{\rho_{E E^\prime}^\textnormal{drift}-\rho_{E E^\prime}^\textnormal{ideal}} \to 0$ as $m \to \infty$. Therefore, the covertness analysis remains unchanged.
\end{proof}

\begin{figure}[!htbp]
    \centering
    \includegraphics[width=0.48\textwidth]{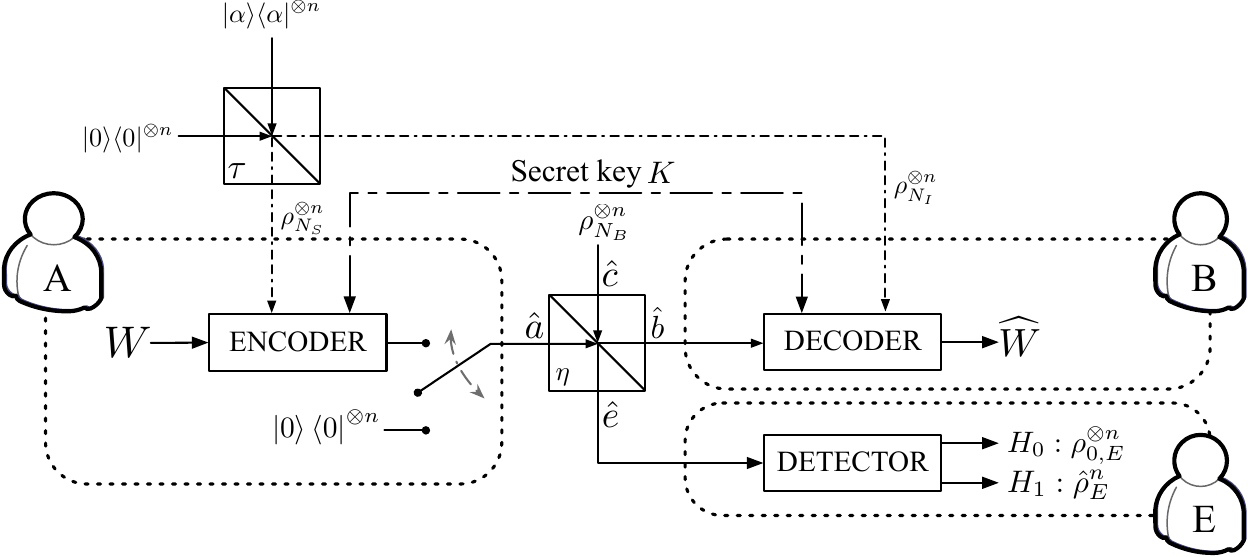}
    \caption{One-way covert communication model with coherent ring states.}
    \label{fig:one_way_covert_coherent}
\end{figure}

To demonstrate again the versatility of ring states for covert communication, we also consider a one-way covert communication model based on coherent states rather than \ac{TMSV} states~\cite{Bash2015QuantumsecureCovert,Gagatsos2020CovertCapacity,Wang2022CharacterizationCovert}. In the model shown in Fig.~\ref{fig:one_way_covert_coherent}, a coherent state is sent through a beamsplitter with transmissivity $\tau$, and the two output modes are distributed to Alice and Bob, respectively, which captures for instance the distribution of a phase reference.

\begin{theorem}
\label{thm:coherent_one_way_covert}
There exist $\varepsilon$-reliable and $\delta$-covert codes operating on $n$ phase-modulated coherent states without secret keys such that
\begin{multline*}
	\lim_{n \rightarrow \infty} \frac{\log M}{\sqrt{n}} 
\geq
	\frac{2 \eta}{1-\eta}
	\sqrt{\frac{\eta}{1-\eta}N_T \left( \frac{\eta}{1-\eta}N_T+1 \right)} \\
\times
	\log \left(1+\frac{1}{N_T} \right)\; \delta.
\end{multline*}
\end{theorem}
\begin{proof}[Sketch of Proof]
We follow steps similar to those in Theorem~\ref{thm:TMSV_one_way_covert}. As shown in Appendix~\ref{sec:coherent_one_way_covert_proof}, the covertness constraint reduces to the trace distance $\frac{1}{2}\norm[1]{\smash{{\rho}_{n,E}^{\otimes n}-\rho_{0,E}^{\otimes n}}}$, which is less than $\delta$ if $N_S \leq \frac{1}{\sqrt{n}}\frac{\delta}{\sqrt{C_2^\prime}}$, where $C_2^\prime = \frac{(1-\eta)^2}{4\frac{\eta}{1-\eta}N_T \left( \frac{\eta}{1-\eta}N_T +1 \right)}$. The result of Theorem~\ref{thm:coherent_one_way_covert} is obtained by evaluating the leading term of Proposition~\ref{prop:coherent_low_power_one_way} at the maximum permissible signal power $N_S = \frac{1}{\sqrt{n}}\frac{\delta}{\sqrt{C_2^\prime}}$.
\end{proof}

The result in Theorem~\ref{thm:coherent_one_way_covert} coincides with the previous result~\cite[Proposition 3.2]{Wang2022CharacterizationCovert}, for which the modulation format is restricted to \ac{QPSK} modulation. However, the constraint on $N_S$ in the proof of Theorem~\ref{thm:coherent_one_way_covert} is derived via a rather direct fidelity-based analysis of ring states, instead of a lengthy Taylor expansion of the modulated states under all possible phase shifts.

\subsection{Round-Trip Covert Communication}
\label{sec:round_trip_covert}

\begin{figure}[!htbp]
    \centering
    \includegraphics[width=0.84\linewidth]{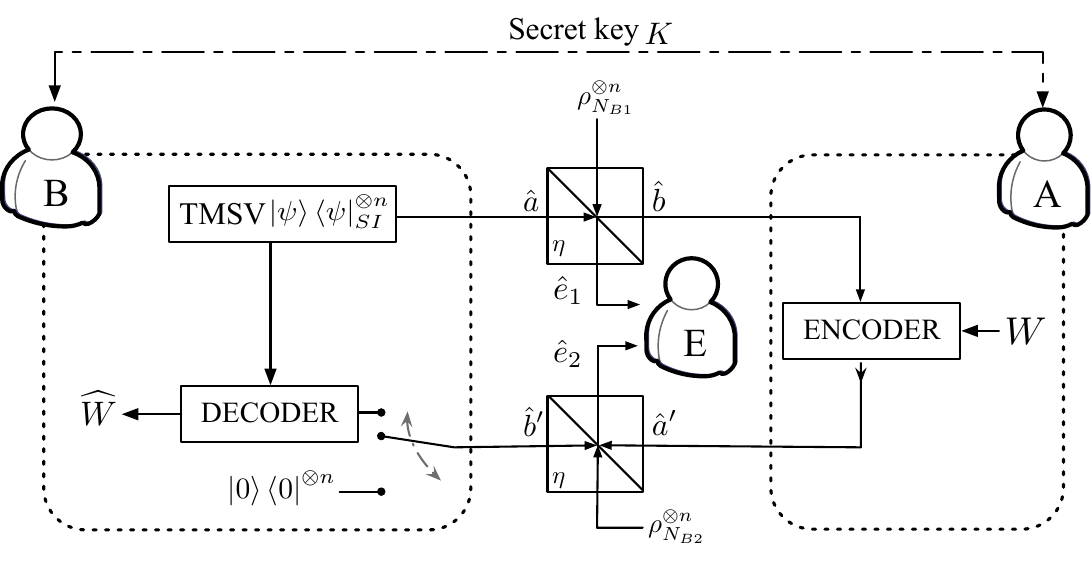}
    \caption{Round-trip covert communication model with TMSV ring states.}
    \label{fig:round_trip_covert}
\end{figure}

We finally consider round-trip covert communication using \ac{TMSV} ring states, as illustrated in Fig.~\ref{fig:round_trip_covert}. This model extends the round-trip framework introduced in Sec.~\ref{sec:comm-comm} by incorporating an eavesdropper (Eve) who monitors the forward and backward channels via systems $E_1$ and $E_2$, with corresponding annihilation operators $\hat{e}_1$ and $\hat{e}_2$, while remaining oblivious to the presence of the communication.

The covertness constraint of the round-trip model is given by
$\frac{1}{2}\norm[1]{\smash{\hat{\rho}_{E_1E_2}^n - {\rho}_{0,E_1E_2}^{\otimes n}}} \leq \delta$, where
$\hat{\rho}_{E_1E_2}^n = \frac{1}{M L}\sum_{w=1}^M \sum_{k=1}^L \mathcal{E}_{A \rightarrow E_1 E_2}^{\otimes n} \; \rho_{A^n}^{(w,k)}$ denotes the average state received by Eve when communication occurs, and $\mathcal{E}_{A \rightarrow E_1 E_2}$ represents the lossy thermal bosonic channel from Alice's system $A$ to Eve's system $E_1 E_2$. We define the no-communication state ${\rho}_{0,{E_1 E_2}}$ as the state received by Eve when $N_S=0$ and the phase difference between $E_1$ and $E_2$ is uniformly distributed over $[0,2\pi)$~\footnote{This uniform phase randomization eliminates any residual codeword structure in Eve's state, thereby rendering it similar to thermal noise.}.

\begin{theorem}
\label{thm:TMSV_roundtrip_covert}
In the round-trip model, there exist $\varepsilon$-reliable and $\delta$-covert codes operating on $n$ phase-modulated \ac{TMSV} states with the assistance of a secret key such that
\begin{multline*}
	\lim_{n \rightarrow \infty} \frac{\log M}{\sqrt{n}\log n} \\
\geq
	\frac{\eta^2}{2 \Bigl((1-\eta)\eta N_{B1}+(1-\eta)N_{B2}+1 \Bigr)}
	\frac{1}{\sqrt{C_2^{\prime \prime}}} \; (\delta-\kappa),
\end{multline*}
where $C_2^{\prime \prime}$ is given by~\eqref{eq:C2_prime_prime} in Appendix~\ref{sec:TMSV_roundtrip_covert_proof}.
\end{theorem}
\begin{proof}[Sketch of Proof]
The covertness constraint for the round-trip model can be upper bounded as
\begin{multline*}
	\frac{1}{2}\norm[1]{
	\hat{\rho}_{E_1E_2}^n - {\rho}_{0,E_1E_2}^{\otimes n}
	} \\
	\leq
	\frac{1}{2}\norm[1]{
	\hat{\rho}_{E_1 E_2}^n-\rho_{n,E_1 E_2}^{\otimes n}
	}
	+
	\frac{1}{2}\norm[1]{
	\rho_{n,E_1 E_2}^{\otimes n}-\rho_{0,E_1 E_2}^{\otimes n}
	}.
\end{multline*}
As shown in Appendix~\ref{sec:TMSV_roundtrip_covert_proof}, one can bound the first term $\frac{1}{2}\norm[1]{
	\smash{\hat{\rho}_{E_1 E_2}^n-\rho_{n,E_1 E_2}^{\otimes n}}
	} \leq \kappa$ for $\kappa \in (0,\frac{\delta}{2})$ using an information-theoretic argument with secret key consumption.
The second term
$\frac{1}{2}\norm[1]{
	\smash{\rho_{n,E_1 E_2}^{\otimes n}-\rho_{0,E_1 E_2}^{\otimes n}}
	}$
can be upper bounded by $\delta - \kappa$ if $N_S \leq \frac{1}{\sqrt{n}}\frac{\delta-\kappa}{\sqrt{C_2^{\prime \prime}}}$.
The result of Theorem~\ref{thm:TMSV_roundtrip_covert} is then obtained by evaluating the leading term of Proposition~\ref{prop:TMSV_low_power_roundtrip} at the maximum permissible signal power $N_S = \frac{1}{\sqrt{n}}\frac{\delta-\kappa}{\sqrt{C_2^{\prime \prime}}}$. The complete derivation is provided in Appendix~\ref{sec:TMSV_roundtrip_covert_proof}.
\end{proof}

Compared to prior work~\cite{DiCandia2021TwowayCovert}, our round-trip covert throughput remains valid over a broad range of physical parameters, rather than relying on the limiting assumption $\eta \to 0$. Moreover, the number of $\delta$-covert message nats scales as $\sqrt{n}\log n$ in Theorem~\ref{thm:TMSV_roundtrip_covert}, compared to the $\sqrt{n}$ scaling achieved in~\cite{DiCandia2021TwowayCovert}.

\section{Conclusion}
We have analyzed quantum ring states and their applications to non-covert and covert communication over bosonic channels. The diagonal structure of ring states in the Fock basis substantially simplifies the evaluation of information-theoretic quantities such as the von Neumann entropy, despite their non-Gaussian nature.
In particular, we have derived tight and surprisingly simple bounds that characterize achievable communication rates without relying on Taylor expansions, and we have designed covert communication protocols that are more efficient than those previously reported.
The usefulness of quantum ring states also extends beyond the present work as seen in recent applications to advanced coding schemes for covert communication~\cite{Wang2024ResourceefficientEntanglementassisted} and joint communication-sensing models~\cite{Erdogan2026JointCommunication}, pointing to several promising directions for other applications.

\clearpage
\onecolumn
\appendix

\counterwithin{lemma}{section}
\renewcommand{\thelemma}{\Alph{section}\arabic{lemma}}

\section{Proof of Proposition~\ref{prop:TMSV_ring_converge} \\
(Convergence speed of the \texorpdfstring{$m$}{m}-ary PSK joint state to the TMSV ring state)}
\label{sec:TMSV_ring_converge_proof}

When $\rho_{SI}$ is a \ac{TMSV} state, the quantum state $\rho_{BI}^{\theta}$ is a zero-mean Gaussian state with a $4 \times 4$ covariance matrix in $(\hat{q}_B, \hat{p}_B, \hat{q}_I, \hat{p}_I)$ form given by~\cite[(9)]{Shi2020PracticalRoute}:
\begin{align}
\label{eq:Bob_cov_one_way}
\boldsymbol{\Lambda}_{BBII,\theta}=
	\begin{bmatrix*}[l]
    	B_{\text{Received}} \matI			& C_{\text{Bob}} \mathbf{R}_\theta\\
    	C_{\text{Bob}} \mathbf{R}_\theta	& S_{\text{Idler}} \matI\\
	\end{bmatrix*},
\end{align}
where $N_T\triangleq N_{B}(1-\eta)$, $B_{\text{Received}}\triangleq 2(N_T+\eta N_S)+1$, $C_{\text{Bob}}\triangleq 2\sqrt{\eta N_S(N_S+1)}$, $S_{\text{Idler}}=2N_S+1$, and $\mathbf{R}_\theta \triangleq
\left[\begin{array}{cc}
	\cos \theta & \sin \theta \\
	\sin \theta & -\cos \theta
\end{array}\right]$.
By first expressing the state in terms of the covariance matrix in $(\hat{a}_B, \hat{a}_I, \hat{a}^\dagger_B, \hat{a}^\dagger_I)$ form and then deriving the p-representation~\cite[Chapter 4.3]{Serafini2023QuantumContinuous}, the density operator $\rho_{BI}^{\theta}$ can be represented in the Fock basis. Setting $\theta=0$, we obtain
{
\begin{align}
\label{eq:rho_BI_theta_zero}
	\rho_{BI}^{\theta = 0}
=
	\sum_{n_1=0}^\infty
	\sum_{n_2=0}^\infty
	\sum_{\Delta=0}^{\min\{n_1,n_2\}}
	\lambda_{n_1, n_2,\Delta}
	\ket{n_1,n_2}
	\bra{n_1-\Delta,n_2-\Delta},
\end{align}
}
where
\begin{align}
	\lambda_{n_1, n_2,\Delta}
&=
		\sqrt{\frac{n_1!n_2!}{(n_1-\Delta)!(n_2-\Delta)!}}
  	 	\left(\frac{N_T - \eta}{N_T - \eta + 1}\right)^{n_1}
    	\left(\frac{N_S\left(N_T-\eta\right)}{N_T\left(N_S+1\right)}\right)^{n_2}
    	\left(\frac{\sqrt{\eta} \sqrt{N_S\left(N_S+1 \right)}}{N_S\left(N_T-\eta\right)}\right)^{\Delta}
    	\nonumber\\
&\quad\times
  		\left(\frac{N_T-\eta}
  				   {\left( N_T-\eta+1 \right) N_T \left( N_S+1 \right)}\right)
  		F_R\left[n_1+1,n_2+1;\Delta +1, \frac{\eta}{\left(N_T-\eta +1\right) N_T} \right],
  		\label{eq:p_n1_n2_n1b_n2b_v1}
\end{align}
which is analogous to~\cite[(B4)]{Shi2020PracticalRoute} up to variable substitution.
Letting $\bar{n}_1=n_1-\Delta$ and $\bar{n}_2=n_2-\Delta$, we obtain an alternative expression
\begin{align}
	\lambda_{\bar{n}_1, \bar{n}_2,\Delta}
  	&=
		\sqrt{\frac{\left(\bar{n}_1+\Delta\right)!
		\left(\bar{n}_2+\Delta\right)!}{\bar{n}_{1}!\bar{n}_{2}!}}
		\left(\frac{N_T}{N_T+1}\right)^{\bar{n}_1}
		\left(\frac{N_S\left(N_T-\eta+1\right)}{\left(N_S+1\right)\left(N_T+1\right)}\right)^{\bar{n}_2}
		\left(\frac{\sqrt{\eta}}{\left(N_T+1\right)} \sqrt{\frac{N_S}{N_S+1}}\right)^{\Delta}
    	\nonumber\\
  	&\quad\times
  		\left(\frac{1}{\left(N_T+1\right)
  		\left(N_S+1\right)}\right)
  		F_R\left[-\bar{n}_1,-\bar{n}_2 ; \Delta+1, \frac{\eta}{\left(N_T-\eta+1\right) N_T}\right]
  		\label{eq:p_n1_n2_n1b_n2b_v2},
\end{align}
where $F_R$ denotes the regularized hypergeometric function. The alternative expression will prove useful when negative binomials are involved.

To obtain the Fock coefficients of ${\rho}_{BI}^\textnormal{TMSV-ring}$, we let
\begin{align*}
	{\rho}_{BI,m}^\textnormal{TMSV}
\triangleq
	\frac{1}{m}
	\sum_{k=0}^{m-1}
	{\rho}_{BI}^{\theta=\frac{2\pi k}{m}}
=
	\frac{1}{m}\sum_{k=0}^{m-1}
	\sum_{n_1=0}^\infty
	\sum_{n_2=0}^\infty
	\sum_{\Delta=0}^{\min\{n_1,n_2\}}
	e^{j \frac{2\pi k}{m} \Delta}
	\lambda_{n_1, n_2,\Delta}
	\ket{n_1,n_2}
	\bra{n_1-\Delta,n_2-\Delta},
\end{align*}
where ${\rho}_{BI}^{\theta=0}$ is defined in~\eqref{eq:rho_BI_theta_zero}, and define
\begin{align*}
{\rho}_{BI}^\textnormal{TMSV-ring}
&\triangleq
	\lim_{m \rightarrow \infty}
	{\rho}_{BI,m}^\textnormal{TMSV}\\
&=
	\lim_{m \rightarrow \infty}
	\frac{1}{m}
	\sum_{k=0}^{m-1}
	\sum_{n_1=0}^\infty
	\sum_{n_2=0}^\infty
	\sum_{\Delta=0}^{\min\{n_1,n_2\}}
	e^{j \frac{2\pi k}{m} \Delta}
	\lambda_{n_1, n_2,\Delta}
	\ket{n_1,n_2}
	\bra{n_1-\Delta,n_2-\Delta} \\
&=
	\sum_{n_1=0}^\infty
	\sum_{n_2=0}^\infty
	\delta(\Delta)
	\lambda_{n_1, n_2,\Delta}
	\ket{n_1,n_2}
	\bra{n_1-\Delta,n_2-\Delta} \\
&=
	\sum_{n_1=0}^\infty
	\sum_{n_2=0}^\infty
	\lambda_{n_1, n_2,\Delta = 0}
	\ket{n_1,n_2}
	\bra{n_1,n_2}.
\end{align*}

We also define
\begin{align}
	p(n_1,n_2) \triangleq \lambda_{n_1,n_2,\Delta=0}.
	\label{eq:p_n1_n2}
\end{align}

By applying the perturbation theorem for the root fidelity~\cite[Theorem 5]{Grace2022PerturbationTheory}, and letting $f(\bar{n}_1, \bar{n}_2, m \Delta)\triangleq\braket{\bar{n}_1+ m \Delta,\bar{n}_2+ m \Delta\vert\rho_{BI}^{\theta=0}}{\bar{n}_1,\bar{n}_2}$, $\lambda_{\bar{n}_1, \bar{n}_2}=\lambda_{\bar{n}_1, \bar{n}_2,\Delta=0}$ in~\eqref{eq:p_n1_n2_n1b_n2b_v2}, and $\nu_m\triangleq\rho_{BI,m}^\textnormal{TMSV}-{\rho}_{BI}^\textnormal{TMSV-ring}$, we obtain
\begin{align}
\label{eq:pert-F-TMSV}
    	1-F(\rho_{BI,m}^\textnormal{TMSV}, {\rho}_{BI}^\textnormal{TMSV-ring})
    &=	\frac{1}{2}
    	\sum_{\Delta=1}^{\infty} \sum_{\bar{n}_1=0}^{\infty}\sum_{\bar{n}_2=0}^{\infty} 
    	\frac{\abs{f(\bar{n}_1, \bar{n}_2, m\Delta)}^2}
    	{\lambda_{\bar{n}_1, \bar{n}_2}+\lambda_{\bar{n}_1+m\Delta, \bar{n}_2+m\Delta}}
    	+O(\norm[2]{\nu_m}^3).
\end{align}

We introduce the following variables to simplify the expressions:
$a\triangleq \frac{N_T} {N_T+1}$,
$b\triangleq \frac{ N_S \left( N_T-\eta +1 \right) }{ \left( N_S+1 \right) \left( N_T+1 \right)  }$,
$c\triangleq \frac{\sqrt{\eta} }{ (N_T+1)} \sqrt{\frac{N_S}{N_S+1}}$,
$d\triangleq \frac{1}{(N_T+1)(N_S+1)}$, and
$z\triangleq \frac{\eta}{\left(N_T-\eta+1\right) N_T}$.
We assume $0<z<1$, leading to the constraint $N_T > \eta$.

To evaluate~\eqref{eq:pert-F-TMSV}, we first consider the case $m=1$ and omit the big-O term in~\eqref{eq:pert-F-TMSV}. We have
\begin{flalign}
	&   \;\quad
		\frac{1}{2}
		\sum_{\Delta=1}^{\infty} \sum_{\bar{n}_1=0}^{\infty}\sum_{\bar{n}_2=0}^{\infty}
		\frac{\abs{f(\bar{n}_1, \bar{n}_2, \Delta)}^2}
			 {\lambda_{\bar{n}_1, \bar{n}_2}+\lambda_{\bar{n}_1+\Delta, \bar{n}_2+\Delta}}&\nonumber\\
    &=
    	\frac{1}{2}
    	\sum_{\Delta=1}^{\infty} \sum_{\bar{n}_1=0}^{\infty}\sum_{\bar{n}_2=0}^{\infty} 
		\frac{ \left| \left\langle\bar{n}_1+\Delta, \bar{n}_2+\Delta| \rho_{BI}^{\theta = 0} | \bar{n}_1, \bar{n}_2\right\rangle \right|^2 }
               { \left\langle\bar{n}_1+\Delta, \bar{n}_2+\Delta| \rho_{BI}^{\theta = 0} | \bar{n}_1+\Delta, \bar{n}_2+\Delta \right\rangle
                 \; + \; \left\langle\bar{n}_1, \bar{n}_2| \rho_{BI}^{\theta = 0} | \bar{n}_1, \bar{n}_2\right\rangle  } \nonumber\\
    &= 	\frac{1}{2}
    	\sum_{\Delta=1}^{\infty} \sum_{\bar{n}_1=0}^{\infty} \sum_{\bar{n}_2=0}^{\infty}
        \binom{\bar{n}_1+\Delta}{\Delta} \binom{\bar{n}_2+\Delta}{\Delta}
        \left( a^{\bar{n}_1} b^{\bar{n}_2} c^{2 \Delta} d  \right)
        \times {}_2F_1 \left[-\bar{n}_1 ,-\bar{n}_2; \Delta +1, z \right] \nonumber\\
    &   \quad \times
        \frac{ {}_2F_1 \left[-\bar{n}_1 ,-\bar{n}_2; \Delta +1, z \right]}
              { {}_2F_1 \left[-\bar{n}_1 ,-\bar{n}_2; 1, z \right] + \left( ab \right)^\Delta \times {}_2F_1 \left[-\bar{n}_1 -\Delta,-\bar{n}_2-\Delta ; 1, z \right]} \nonumber\\
    & 	\leq 
    	\frac{1}{2}
    	\sum_{\Delta=1}^{\infty} \sum_{\bar{n}_1=0}^{\infty} \sum_{\bar{n}_2=0}^{\infty}
        \binom{\bar{n}_1+\Delta}{\Delta} \binom{\bar{n}_2+\Delta}{\Delta}
        \left( a^{\bar{n}_1} b^{\bar{n}_2} c^{2 \Delta} d  \right)
        \times {}_2F_1 \left[-\bar{n}_1 ,-\bar{n}_2; \Delta +1, z \right]
        \label{eq:Upsilon-for-TMSV-f-uni} \\
    &=	\frac{1}{2}
    	\sum_{\Delta=1}^{\infty} \sum_{\bar{n}_1=0}^{\infty} \sum_{\bar{n}_2=0}^{\infty}
        \binom{\bar{n}_1+\Delta}{\Delta} \binom{\bar{n}_2+\Delta}{\Delta}
        \left( a^{\bar{n}_1} b^{\bar{n}_2} c^{2 \Delta} d  \right)
        \times \sum_{k=0}^{\min \left(\bar{n}_1, \bar{n}_2\right)} \frac{\left(-\bar{n}_1\right)_k\left(-\bar{n}_2\right)_k}{(\Delta+1)_k}
        \left( \frac{z^k}{k !} \right) \nonumber\\
    &=  \frac{1}{2}
    	\sum_{\Delta=1}^{\infty} \sum_{\bar{n}_1=0}^{\infty} \sum_{\bar{n}_2=0}^{\infty}
        \left( a^{\bar{n}_1} b^{\bar{n}_2} c^{2 \Delta} d  \right) 
        \times
        \sum_{k=0}^{\min \left(\bar{n}_1, \bar{n}_2\right)}
        \frac{ \left( \bar{n}_1 + \Delta \right) !}{ \left( \bar{n}_1 - k \right) !}
        \frac{ \left( \bar{n}_2 + \Delta \right) !}{ \left( \bar{n}_2 - k \right) !}
        \frac{1}   { \Delta ! \left( \Delta + k \right)! }
        \left( \frac{z^k}{k !} \right) \nonumber\\
    & =	\frac{1}{2}
    	\sum_{\Delta=1}^{\infty} \sum_{k=0}^{\infty} \sum_{u_1=0}^{\infty} \sum_{u_2=0}^{\infty}
        \left( a^{u_1 + k} b^{u_2 + k} c^{2 \Delta} d  \right) 
        \frac{ \left( u_1 + \Delta + k \right) !}{ u_1 !}
        \frac{ \left( u_2 + \Delta + k \right) !}{ u_2 !}
        \frac{1}   { \Delta ! \left( \Delta + k \right)! }
        \left( \frac{z^k}{k !} \right)
        \label{eq:U_replace}\\
    &=	\label{eq:n_binom_1}
        \frac{1}{2}
        \sum_{\Delta=1}^{\infty} \sum_{k=0}^{\infty} \left( a b \right)^k c^{2 \Delta} d 
        \left[ \left( 1-a \right)^{-\Delta -k -1} \right]
        \left[ \left( 1-b \right)^{-\Delta -k -1} \right]
        \frac{ \left( \Delta + k \right) !}   { \Delta ! k !} z^k \\
    &=  \frac{1}{2}
    	\sum_{\Delta=1}^{\infty} c^{2 \Delta} d
        \left( 1-a \right)^{-\Delta -1}
        \left( 1-b \right)^{-\Delta -1}
        \sum_{k=0}^{\infty} \binom{k + \Delta}{k}
        \left[ \left( \frac{a}{1-a} \right) \left( \frac{b}{1-b} \right) z \right]^k \nonumber\\
    &=	\label{eq:Is_one}
        \frac{1}{2}
        \sum_{\Delta=1}^{\infty}
        \left( \frac{\eta N_S}{N_T+1} \right)^\Delta
\end{flalign}
where~\eqref{eq:Upsilon-for-TMSV-f-uni} follows by upper bounding the last term on the previous line by one,~\eqref{eq:U_replace} follows from the substitutions $u_1 = \bar{n}_1 -k $ and $u_2 = \bar{n}_2 -k $,~\eqref{eq:n_binom_1} and~\eqref{eq:Is_one} follow by summing the negative binomial series, substituting back the parameters $a$, $b$, $c$, $d$, and $z$, and assuming $\left( \frac{\eta N_S}{N_T+1} \right) < 1$.\\

We then extend the result to $m$-ary \ac{PSK} and obtain
\begin{align*}
     \frac{1}{2}
     \sum_{\Delta=1}^{\infty} \sum_{\bar{n}_1=0}^{\infty}\sum_{\bar{n}_2=0}^{\infty}
     \frac{\abs{f(\bar{n}_1, \bar{n}_2, m\Delta)}^2}
     	  {\lambda_{\bar{n}_1, \bar{n}_2}+\lambda_{\bar{n}_1+m\Delta, \bar{n}_2+m\Delta}}
\leq 
	\frac{1}{2}
	\sum_{\Delta=1}^{\infty}
    \left( \frac{\eta N_S}{N_T+1} \right)^{m\Delta}
    =\frac{\frac{1}{2} \left( \frac{\eta N_S}{N_T+1} \right)^{m}}
          {1- \left( \frac{\eta N_S}{N_T+1} \right)^{m}},     
\end{align*}
where the equality follows by assuming $\left( \frac{\eta N_S}{N_T+1} \right) < 1$.

For the big-O term in~\eqref{eq:pert-F-TMSV}, we analyze the Hilbert-Schmidt norm of $\nu_m$ and verify that it vanishes as $m \to \infty$. Note that $\{\ket{n_1,n_2}_{BI}\}_{n_1,n_2\in\bbN}$ forms an orthonormal basis for $\calH_B\otimes\calH_I$. By definition of $\norm[2]{\cdot}$,
\begin{align}
  	\norm[2]{\nu_m}^2
&\triangleq
	\sum_{{n}_1=0}^{\infty}\sum_{{n}_2=0}^{\infty}
	\sum_{\bar{n}_1=0}^{\infty}\sum_{\bar{n}_2=0}^{\infty}
	\abs{\bra{n_1,n_2} \nu_{m} \ket{\bar{n}_1,\bar{n}_2}}^2 \nonumber\\
&=
	2
	\sum_{\Delta=1}^{\infty} \sum_{\bar{n}_1=0}^{\infty}\sum_{\bar{n}_2=0}^{\infty}
	\abs{f(\bar{n}_1, \bar{n}_2, m\Delta)}^2
	\nonumber\\
&=
	2
	\sum_{\Delta=1}^{\infty} \sum_{\bar{n}_1=0}^{\infty}\sum_{\bar{n}_2=0}^{\infty}
	\binom{\bar{n}_1+m\Delta}{m\Delta}
	\binom{\bar{n}_2+m\Delta}{m\Delta}
	a^{2\bar{n}_1}b^{2\bar{n}_2}c^{2m\Delta}d^2
	{}_2F_1[-\bar{n}_1, -\bar{n}_2; m\Delta+1, z]^2
	\nonumber\\
&=
	2
	\sum_{\Delta=1}^{\infty} \sum_{\bar{n}_1=0}^{\infty}\sum_{\bar{n}_2=0}^{\infty}
	\binom{\bar{n}_1+m\Delta}{m\Delta}
	\binom{\bar{n}_2+m\Delta}{m\Delta}
	a^{\bar{n}_1}
	b^{\bar{n}_2}
	c^{2m\Delta}
	d \:
	{}_2F_1[-\bar{n}_1, -\bar{n}_2; m\Delta+1, z]
	\nonumber\\
&	\quad
	\times
	a^{\bar{n}_1}
	b^{\bar{n}_2}
	d \:
	{}_2F_1[-\bar{n}_1, -\bar{n}_2; m\Delta+1, z]
	\nonumber\\
&\leq
	2
	\sum_{\Delta=1}^{\infty}
	\left(\frac{\eta N_S}{N_T+1}\right)^{m\Delta}
	\label{eq:prop1_last}
	\\
&=	\frac{2 \left(\frac{\eta N_S}{N_T+1}\right)^m}
         {1-\left(\frac{\eta N_S}{N_T+1}\right)^m}
    \nonumber,
\end{align}
where~\eqref{eq:prop1_last} follows by upper bounding $a^{\bar{n}_1}
	b^{\bar{n}_2}
	d \:
	{}_2F_1[-\bar{n}_1, -\bar{n}_2; m\Delta+1, z]$ on the previous line by one and reusing the derivation from~\eqref{eq:Upsilon-for-TMSV-f-uni} to~\eqref{eq:Is_one}. With $N_T>\max \left\{\eta, \eta N_S-1\right\}$, we obtain $\left\|\nu_m\right\|_2^2=O\left(\left(\frac{\eta N_S}{N_T+1}\right)^m\right)$ and $\left\|\nu_m\right\|_2^3=O\left(\left(\frac{\eta N_S}{N_T+1}\right)^\frac{3m}{2}\right)$ as $ m \rightarrow \infty$.

\section{Proof of Proposition~\ref{prop:coherent_ring_converge} \\
(Convergence speed of the \texorpdfstring{$m$}{m}-ary PSK joint state to the coherent ring state)}
\label{sec:coherent_ring_converge_proof}

In the one-way communication model with coherent ring states depicted in Fig.~\ref{fig:one_way_covert_coherent}, Bob's covariance matrix for the received and idler modes in $(\hat{q}_R, \hat{p}_R, \hat{q}_I, \hat{p}_I)$ form is given by
{
\begin{align*}
\boldsymbol{\Lambda}_{\theta, RRII}^{\prime \prime}=
  \begin{bmatrix}
	1+ 2N_T	&0			&0	&0
\\
	0		&1+ 2N_T	&0	&0
\\
	0		&0			&1	&0
\\
	0		&0			&0	&1
\end{bmatrix} = \left[ \left( 1+2N_T \right) I_2 \right] \oplus I_2.
\end{align*}
}
For Gaussian states, a block-diagonal (direct-sum) covariance matrix indicates that the two modes are in a tensor-product (uncorrelated) state. The corresponding mean vector is given by
{
\begin{align*}
	\bar{r}_{\text{Bob,qpqp}}
=
\begin{bmatrix}
	\sqrt{2} \operatorname{Re}\left\{\sqrt{\eta \tau} \alpha e^{j \theta}\right\}
	\\
	\sqrt{2} \operatorname{Im}\left\{\sqrt{\eta \tau} \alpha e^{j \theta}\right\}
	\\
	\sqrt{2} \operatorname{Re}\left\{\sqrt{1- \tau} \alpha \right\}
	\\
	\sqrt{2} \operatorname{Im}\left\{\sqrt{1- \tau} \alpha \right\}
\end{bmatrix}.
\end{align*}}

Based on the covariance matrix and the mean vector, we derive Bob's Fock coefficients
\begin{align}
&
	\bra{n} \rho_B^{\theta=0} \ket{n-\Delta} \nonumber\\
&= 
	p(n,\Delta) \nonumber\\
&\triangleq
	\sqrt{\frac{n!}{(n-\Delta)!}}
	\frac{e^{\left(-\frac{\eta \tau \abs{\alpha}^2}{N_T}\right)}}{(N_T+1)\Delta!}
	\left(\frac{N_T}{N_T+1} \right)^n
	\left( \frac{\sqrt{\eta \tau} \abs{\alpha}}{N_T}\right)^\Delta
	{}_1 F_1\left[n+1; \Delta+1 ; \frac{\eta \tau \abs{\alpha}^2}{N_T(N_T+1)} \right]
	\label{eq:p_n_Delta},
\end{align}
where ${ }_1 F_1[p ; q ; z]\triangleq\sum_{k=0}^{\infty}\frac{(p)_k \, z^k}{(q)_k \, k!}$, and $(p)_k$ is the Pochhammer symbol.

By applying the perturbation theorem for the root fidelity~\cite[Theorem 5]{Grace2022PerturbationTheory}, and letting $f(n, m \Delta)\triangleq \braket{n\vert\rho_{B}^{\theta=0}}{n-m \Delta}$, $\lambda_{n}=p(n,\Delta=0)$ in~\eqref{eq:p_n_Delta}, and $\nu_m\triangleq\rho_{B,m}^\textnormal{coherent}-{\rho}_{B}^\textnormal{coherent-ring}$, we obtain
\begin{align}
\label{eq:pert-F-coherent}
    	1-F(\rho_{B,m}^\textnormal{coherent}, {\rho}_{B}^\textnormal{coherent-ring})
    &=	\frac{1}{2}
    	\sum_{\Delta=1}^{\infty} \sum_{n=0}^{\infty} 
    	\frac{\abs{f(n, m\Delta)}^2}
    	{\lambda_{n}+\lambda_{n-m\Delta}}
    	+O(\norm[2]{\nu_m}^3).
\end{align}

We introduce the following variables to simplify the expressions:
$a\triangleq \frac{N_T} {N_T+1}$,
$c\triangleq \frac{\sqrt{\eta \tau }\abs{\alpha}}{N_T}$,
$d\triangleq \frac{e^{\left(-\frac{\eta \tau \abs{\alpha}^2}{N_T}\right)}}{(N_T+1)}$, and
$z\triangleq \frac{\eta \tau \abs{\alpha}^2}{N_T(N_T+1)}$.
To evaluate~\eqref{eq:pert-F-coherent}, we first consider the case $m=1$ and omitting the big-O term in~\eqref{eq:pert-F-coherent}. We have
\begin{flalign}
&\;\quad
	\frac{1}{2}
	\sum_{\Delta=1}^{\infty} \sum_{n=0}^{\infty} 
	\frac{\abs{f(n, \Delta)}^2}
    	 {\lambda_{n}+\lambda_{n-\Delta}} &\nonumber\\
&=
	\frac{1}{2}
	\sum_{\Delta=1}^{\infty} \sum_{n=\Delta}^{\infty} 
	\frac{\abs{ \braket{n\vert\rho_{BI}^{\theta=0}}{n-\Delta} }^2}
    	 {\braket{n\vert\rho_{BI}^{\theta=0}}{n}
	      +\braket{n-\Delta \vert\rho_{BI}^{\theta=0}}{n-\Delta}} \nonumber \\
&=
	\frac{1}{2}
	\sum_{\Delta=1}^{\infty} \sum_{n=\Delta}^{\infty}
	\binom{n}{\Delta} \frac{1}{\Delta !} a^n c^{2 \Delta} d \;
	{}_1 F_1\left[n+1; \Delta+1 ; z \right]
	\cdot
	\left(
	\frac{{}_1 F_1\left[n+1; \Delta+1 ; z \right]}
    	 {a^{-\Delta} {}_1 F_1\left[n-\Delta+1; 1 ; z \right]
          + {}_1 F_1\left[n+1; 1 ; z \right]}
	\right)
	\nonumber \\
&\leq
	\frac{1}{2}
	\sum_{\Delta=1}^{\infty} \sum_{n=\Delta}^{\infty}
	\binom{n}{\Delta} \frac{1}{\Delta !} a^n c^{2 \Delta} d \;
	{}_1 F_1\left[n+1; \Delta+1 ; z \right]
	\label{eq:coherent_drop} \\
&\leq
	\frac{1}{2}
	\sum_{\Delta=1}^{\infty} \sum_{n=\Delta}^{\infty}
	\binom{n}{\Delta} \frac{1}{\Delta !} a^n c^{2 \Delta} d \;
	\left( 1+\frac{z}{\Delta+1}\right)^n
	e^z
	\label{eq:coherent_bound_2},
\end{flalign}
where~\eqref{eq:coherent_drop} follows from upper bounding the last term on the previous line by one, and~\eqref{eq:coherent_bound_2} follows from bounding ${}_1 F_1\left[p; q ; z \right]$ by $\left( 1+\frac{z}{q}\right)^{p-1}e^z$ for $p \geq q \geq 2$.\\

We then extend the result to $m$-ary \ac{PSK} and obtain
\begin{flalign}
&   \;\quad
		\frac{1}{2}
    	\sum_{\Delta=1}^{\infty} \sum_{n=0}^{\infty} 
    	\frac{\abs{f(n, m\Delta)}^2}
    	{\lambda_{n}+\lambda_{n-m\Delta}} &\nonumber\\
&\leq
    	\frac{1}{2}
    	\sum_{\Delta=1}^{\infty} \sum_{n=m\Delta}^{\infty}
    	\binom{n}{m\Delta} \frac{1}{(m\Delta)!} a^n c^{2 m\Delta} d \;
    	\left( 1+\frac{z}{m\Delta+1}\right)^n
    	e^z \nonumber \\
&=
    	\frac{d e^z}{2}
    	\sum_{\Delta=1}^{\infty}
    	\frac{c^{2 m\Delta}}{(m\Delta)!}
    	\sum_{n=m\Delta}^{\infty}
    	\binom{n}{m\Delta}
		\left(a
    	 	\left( 1+\frac{z}{m\Delta+1}\right)
    	\right)^n
    	\nonumber \\
&\leq
    	\frac{d e^z}{2}
    	\sum_{\Delta=1}^{\infty}
    	\frac{c^{2 m\Delta}}{(m\Delta)!}
    	\frac{1}{ 	\left(1-\left(a
    	 				\left( 1+\frac{z}{2}\right)
    				\right) \right)^{m\Delta+1}}
    	\label{eq:coherent_cov_1} \\
&\leq
    	\frac{d e^z}{2  \left(1-\left(a
    	 				\left( 1+\frac{z}{2}\right)\right) \right)}
    	\sum_{\Delta=1}^{\infty}
    	\frac{1}{(m\Delta)!}
    	\left(
    	\frac{c^2}
    	     {\left(1-\left(a\left( 1+\frac{z}{2}\right)\right)\right)}
    	\right)^{m\Delta} \nonumber \\
&\leq		
		\frac{\left(N_T+1\right)}{2\left(N_T+1\right)-\eta \tau\abs{\alpha}^2} 
		e^{\left(
			\frac{2 \eta \tau\abs{\alpha}^2\left(N_T+1\right)^2}
		     {\left(2\left(N_T+1\right)-\eta \tau \abs{\alpha}^2\right)N_T^2}
		     -\frac{\eta \tau\abs{\alpha}^2}{N_T+1}\right)}
		\frac{1}{m!}
		\left(
		\frac{2 \eta \tau\abs{\alpha}^2\left(N_T+1\right)^2}
		     {\left(2\left(N_T+1\right)-\eta \tau \abs{\alpha}^2\right)N_T^2}
		\right)^m
       	\label{eq:coherent_cov_2}\\
&\leq
		\frac{\left(N_T+1\right)}{2\left(N_T+1\right)-\eta \tau\abs{\alpha}^2} 
		e^{
			\frac{2 \eta \tau\abs{\alpha}^2\left(N_T+1\right)^2}
		         {\left(2\left(N_T+1\right)-\eta \tau \abs{\alpha}^2\right)N_T^2}
		  }
		\frac{1}{m!}
		\left(
		\frac{2 \eta \tau\abs{\alpha}^2\left(N_T+1\right)^2}
		     {\left(2\left(N_T+1\right)-\eta \tau \abs{\alpha}^2\right)N_T^2}
		\right)^m \nonumber,
\end{flalign}
where~\eqref{eq:coherent_cov_1} holds provided $a\left( 1+\frac{z}{2}\right)<1 \Leftrightarrow N_T>\frac{\eta \tau \abs{\alpha}^2}{2}-1$, and~\eqref{eq:coherent_cov_2} follows from $\sum_{\Delta=1}^\infty\frac{u^{m \Delta}}{(m \Delta)!} \leq \frac{u^m}{m!} e^u$ and substituting the variables.

For the big-O term in~\eqref{eq:pert-F-coherent}, we analyze the Hilbert-Schmidt norm of $\nu_m$ and verify that it vanishes as $m \to \infty$. Note that $\{\ket{n}_{B}\}_{n\in\bbN}$ forms an orthonormal basis for $\calH_B$. By the definition of $\norm[2]{\cdot}$,
{
\begin{align}
  	\norm[2]{\nu_m}^2
&\triangleq
	2
    \sum_{\Delta=1}^{\infty} \sum_{n=0}^{\infty} 
    \abs{f(n, m\Delta)}^2
    \nonumber\\
&=
   	2
   	\sum_{\Delta=1}^{\infty} \sum_{n=m\Delta}^{\infty}
   	\binom{n}{m\Delta} \frac{1}{(m\Delta)!} a^{2n} c^{2m\Delta} d^2
   	{}_1 F_1\left[n+1; m\Delta+1 ; z \right]^2
   	\nonumber\\
&\leq
   	2
   	\sum_{\Delta=1}^{\infty} \sum_{n=m\Delta}^{\infty}
   	\binom{n}{m\Delta} \frac{1}{(m\Delta)!} a^{2n} c^{2m\Delta} d^2
   	e^{2z}
   	\left( 1+\frac{z}{\Delta+1}\right)^{2n}
   	\nonumber\\
&\leq
   	\frac{2 d^2 e^{2 z}}
   	     {1-a^2\left(1+\frac{z}{2}\right)^2}
   	e^{\left(\frac{c^2 a^2\left(1+\frac{z}{2}\right)^2}{1-a^2\left(1+\frac{z}{2}\right)^2}\right)}
   	\frac{1}{m!}
   	\left(\frac{c^2 a^2\left(1+\frac{z}{2}\right)^2}{1-a^2\left(1+\frac{z}{2}\right)^2}\right)^m,
    \label{eq:coherent_O}
\end{align}
}where~\eqref{eq:coherent_O} follows by $1+\frac{z}{\Delta+1}\leq 1+\frac{z}{2}$ for $\Delta \geq 1$, $\sum_{n=m \Delta}^{\infty} \binom{n}{m\Delta}  r^n=\frac{r^{m \Delta}}{(1-r)^{m \Delta+1}}$, and similar techniques in~\eqref{eq:coherent_cov_2}.
Substituting the variables and further upper-bounding the resulting expression, we evaluate the expression $O\left(\norm[2]{\nu_m}^3\right)$ as { $O\left( 
\frac{1}{ \left( m! \right)^{\frac{3}{2}}}\left(\frac{2 \eta \tau\abs{\alpha}^2\left(N_T+1\right)^2}{\left(2\left(N_T+1\right)-\eta \tau \abs{\alpha}^2\right) N_T^2}\right)^{\frac{3}{2}m}\right)$}.

\section{Proof of Proposition~\ref{prop:TMSV_m_penalty} \\
(Difference in Holevo information between the TMSV ring state and the \texorpdfstring{$m$}{m}-ary PSK TMSV state)}
\label{sec:TMSV_m_penalty_proof}

By applying the perturbation theorem for the von Neumann entropy~\cite[Theorem 2]{Grace2022PerturbationTheory}, and reusing the functions and variables defined in Appendix~\ref{sec:TMSV_ring_converge_proof}, we obtain
\begin{align}
\label{eq:pert-H}
\vonH{\rho_{BI,m}^\textnormal{TMSV}}-\vonH{{\rho}_{BI}^\textnormal{TMSV-ring}}
&=	-
	\sum_{\Delta=1}^{\infty}\sum_{\bar{n}_1=0}^{\infty}\sum_{\bar{n}_2=0}^{\infty}
	\abs{f(\bar{n}_1, \bar{n}_2, m\Delta)}^2 \;
	\frac{\log(\lambda_{\bar{n}_1+m\Delta, \bar{n}_2+m\Delta}) - \log( \lambda_{\bar{n}_1, \bar{n}_2})}
    	 {\lambda_{\bar{n}_1+m\Delta, \bar{n}_2+m\Delta} - \lambda_{\bar{n}_1, \bar{n}_2}}
    	 \nonumber\\
&\quad
	+O(\norm[2]{\nu_m}^3),
\end{align}
where $\nu_m\triangleq\rho_{BI,m}^\textnormal{TMSV}-{\rho}_{BI}^\textnormal{TMSV-ring}$.

Considering the case $m=1$ and omitting the big-O term in~\eqref{eq:pert-H}, we obtain
\begin{align}
	0
	&<	\sum_{\Delta=1}^{\infty}\sum_{\bar{n}_1=0}^{\infty}\sum_{\bar{n}_2=0}^{\infty}
    	\abs{f(\bar{n}_1, \bar{n}_2, \Delta)}^2 \;
		\frac{\log(\lambda_{\bar{n}_1+\Delta, \bar{n}_2+\Delta}) - \log( \lambda_{\bar{n}_1, \bar{n}_2})}{\lambda_{\bar{n}_1+\Delta, \bar{n}_2+\Delta} - \lambda_{\bar{n}_1, \bar{n}_2}} 
    	\nonumber \\
    &=	 
    	\sum_{\Delta=1}^{\infty}\sum_{\bar{n}_1=0}^{\infty}\sum_{\bar{n}_2=0}^{\infty}
		\binom{\bar{n}_1+\Delta}{\Delta}\binom{\bar{n}_2+\Delta}{\Delta}
		\left(
			a^{2\bar{n}_1}b^{2\bar{n}_2}c^{2 \Delta}d^2
		\right)	
		\left(
			{}_2F_1[-\bar{n}_1,-\bar{n}_2;\Delta+1;z]
		\right)^2 
			\nonumber \\
	&	\quad \times
		\frac{ \Delta \log(ab) + \log\left(\frac{{}_2F_1[-\bar{n}_1-\Delta,-\bar{n}_2-\Delta;1;z]}{{}_2F_1[-\bar{n}_1,-\bar{n}_2;1;z]} \right)}
		     {\left( a^{\bar{n}_1+\Delta}  b^{\bar{n}_2+\Delta}  d \right)  {}_2F_1[-\bar{n}_1-\Delta,-\bar{n}_2-\Delta;1;z] - \left( a^{\bar{n}_1}  b^{\bar{n}_2} d \right)  {}_2F_1[-\bar{n}_1,-\bar{n}_2;1;z]} 
		   	\nonumber \\
	&=	\sum_{\Delta=1}^{\infty}\sum_{\bar{n}_1=0}^{\infty}\sum_{\bar{n}_2=0}^{\infty}
		\binom{\bar{n}_1+\Delta}{\Delta}\binom{\bar{n}_2+\Delta}{\Delta}
		\left(  a^{\bar{n}_1} b^{\bar{n}_2}  c^{2 \Delta}  d \right)
		{}_2F_1[-\bar{n}_1,-\bar{n}_2;\Delta+1;z]
			\nonumber \\
	&	\quad
		\times
		\frac{\Delta \log(ab) + \log(\frac{{}_2F_1[-\bar{n}_1-\Delta,-\bar{n}_2-\Delta;1;z]}{{}_2F_1[-\bar{n}_1,-\bar{n}_2;1;z]})}
		     { \left( a b \right)^{\Delta} \frac{{}_2F_1[-\bar{n}_1-\Delta,-\bar{n}_2-\Delta;1;z]}{{}_2F_1[-\bar{n}_1,-\bar{n}_2;1;z]} -1 }
		\times \frac{{}_2F_1[-\bar{n}_1,-\bar{n}_2;\Delta+1;z]}{{}_2F_1[-\bar{n}_1,-\bar{n}_2;1;z]}
			\nonumber \\
	& < 
		\sum_{\Delta=1}^{\infty}\sum_{\bar{n}_1=0}^{\infty}\sum_{\bar{n}_2=0}^{\infty}
		\binom{\bar{n}_1+\Delta}{\Delta}\binom{\bar{n}_2+\Delta}{\Delta}
		\left(  a^{\bar{n}_1} \; b^{\bar{n}_2} \; c^{2 \Delta} \; d \right)
		{}_2F_1[-\bar{n}_1,-\bar{n}_2;\Delta+1;z] \nonumber \\
	&	\quad
		\times \left( \Delta \left( 2+3z+z \bar{n}_1 + z \bar{n}_2 \right) \right)^2+2.
		\label{eq:H_perturbed_poly}
\end{align}
where~\eqref{eq:H_perturbed_poly} follows by the last term of the previous line $0< \frac{{}_2F_1[-\bar{n}_1,-\bar{n}_2;\Delta+1;z]}{{}_2F_1[-\bar{n}_1,-\bar{n}_2;1;z]} < 1$,
denoting $R=\frac{{ }_2 F_1\left[-\bar{n}_1-\Delta,-\bar{n}_2-\Delta ; 1 ; z\right]}{{ }_2 F_1\left[-\bar{n}_1,-\bar{n}_2 ; 1 ; z\right]}=(1-z)^{2 \Delta} \cdot \frac{{ }_2 F_1\left[1+\bar{n}_1+\Delta, 1+\bar{n}_2+\Delta ; 1 ; z\right]}{{ }_2 F_1\left[1+\bar{n}_1, 1+\bar{n}_2 ; 1 ; z\right]}$,
$\frac{\Delta \log(ab) + \log R}
	  {(ab)^\Delta \; R-1 }
<
		\left(
		\log
		R
		\right)^2+2
$
together with Lemma~\ref{lem:F_ratio_consecutive_delta}, $\log(1+x)<x$, and $0<a,b<1$.

Considering a general $m$, and evaluating~\eqref{eq:H_perturbed_poly} by decomposing the expression into several sums involving negative binomial series, we obtain

\begin{align}
	0
&< 
		\sum_{\Delta=1}^{\infty}\sum_{\bar{n}_1=0}^{\infty}\sum_{\bar{n}_2=0}^{\infty}
		\abs{f(\bar{n}_1, \bar{n}_2, m\Delta)}^2 \;
		\frac{\log(\lambda_{\bar{n}_1+m\Delta, \bar{n}_2+m\Delta}) - \log( \lambda_{\bar{n}_1, \bar{n}_2})}{\lambda_{\bar{n}_1+m\Delta, \bar{n}_2+m\Delta} - \lambda_{\bar{n}_1, \bar{n}_2}} 
    	\nonumber\\
&\leq
		\frac{\phi ^m}{\left(\phi ^m-1\right)^5 N_T^2 \left(-\eta +N_T+1\right)^2}
		\left( P_0+ P_1 N_S + P_2 N_S^2 \right)
		\nonumber\\
&\triangleq
	\zeta(m) \label{eq:zeta_m},
\end{align}
\begin{align*}
\textnormal{ where }
	\phi
&
	\triangleq \frac{\eta N_S}{N_T + 1} <1, \\
P_0 &=
	9 \eta ^2 m^2 \left(\phi ^m-1\right)^2 \left(\phi ^m+1\right)+N_T (N_T (2 \eta ^2 m^4 \left(\phi ^m+1\right) \left(\left(\phi ^m+10\right) \phi ^m+1\right) \\
&\quad
	+2 (\eta -4) \eta  m^3 \left(\left(\phi ^m+4\right) \phi ^m+1\right) \left(\phi ^m-1\right)+4 (3 \eta +1) m^2 \left(\phi ^m+1\right) \left(\phi ^m-1\right)^2\\
&\quad
	+\left(\phi ^m-1\right) N_T (-8 \eta  m^3 \left(\left(\phi ^m+4\right) \phi ^m+1\right)+\left(4 m^2 \left(\phi ^m+1\right)+\left(\phi ^m-1\right)^2\right) \left(\phi ^m-1\right) N_T \\
&\quad
	+8 m^2 \left(\phi ^{2 m}-1\right)+2 (1-\eta ) \left(\phi ^m-1\right)^3)+(1-\eta )^2 \left(\phi ^m-1\right)^4) \\
&\quad	
	-2 \eta  m^2 \left(\phi ^m-1\right) (\eta +(\eta  (7 m-1)-6) \phi ^{2 m}
	+28 \eta  m \phi ^m+7 \eta  m+6)), \\
P_1 &=
	2 \eta ^2 m^2 (N_T((2 \eta +1) m^2 \left(\phi ^m+1\right) \left(\left(\phi ^m+10\right) \phi ^m+1\right) \\
&\quad
	-2 (\eta +3) m \left(\left(\phi ^m+4\right) \phi ^m+1\right) \left(\phi ^m-1\right)-4 \left((m-1) \phi ^{2 m}+4 m \phi ^m+m+1\right) \left(\phi ^m-1\right) N_T\\
&\quad
	+5 \left(\phi ^m+1\right) \left(\phi ^m-1\right)^2)
	-8 \eta  \left(\phi ^m-1\right) \left((m-1) \phi ^{2 m}+4 m \phi ^m+m+1\right)),\\
P_2 &=
	2 \eta ^3 (\eta +1) m^2
	\bigl(11 m^2 \phi ^{2 m}+m^2 \phi ^{3 m}+11 m^2 \phi ^m+m^2\\
&\quad
	-9 m \phi ^{2 m}-2 \phi ^{2 m}-3 m \phi ^{3 m}+2 \phi ^{3 m}+9 m \phi ^m-2 \phi ^m+3 m+2
	\bigr).
\end{align*}

Although the expression in~\eqref{eq:zeta_m} is relatively lengthy, its magnitude is small, as illustrated in Fig.~\ref{fig:TMSV_BPSK_penalty_0d10} for the case of BPSK modulation. Our numerical simulations based on the Walrus library approximate $\chi_{m-\textnormal{PSK}}^\textnormal{TMSV} \approx 2.45\times 10^{-4}$ and $\zeta(m) \approx 6.15\times 10^{-8}$ when $m=2$, $\eta=0.5$, $N_S=10^{-4}$ and $N_T=1$, indicating that the $O(\norm[2]{\nu_m}^3)$ term in~\eqref{eq:pert-H} is negligible in the parameter regime of interest.

\section{Proof of Theorem~\ref{thm:TMSV_ring_Holevo_info} (Holevo information of the TMSV ring state)}
\label{sec:TMSV_ring_Holevo_info_proof}

The Holevo information for this communication model can be broken down into two terms: the von~Neumann entropy of the mixed state under phase modulation and the expected von~Neumann entropy of the conditional state over phases.
\begin{align}
\chi_{\textnormal{ring}}^\textnormal{TMSV}
    &= \vonH{\frac{1}{2 \pi} \int_0^{2 \pi} \rho_{BI}^{\theta} \:  d \theta} 
    	-
    	\left(
    	\frac{1}{2 \pi}\int_0^{2 \pi}
    	\vonH{ \rho_{BI}^{\theta}}
    	\:  d \theta
    	\right)
    	\nonumber \\
    &=
    	\left[
    		-\sum_{n_1=0}^{\infty} \sum_{n_2=0}^{\infty}
    		p\left(n_1, n_2\right) \log p \left(n_1, n_2\right)
    	\right] 
		-
		\left[
			g\left( \frac{\mu_+ - 1}{2}		\right)
		  + g\left( \frac{ \mu_- - 1}{2}  	\right)
  		\right],\label{eq:Holevo_info_cont}
\end{align}
where $g(x) \triangleq (x+1) \log (x+1) - x \log x$ and $\mu_{ \pm}= \sqrt{(N_T+(1+\eta)N_S+1)^2 - 4 \eta N_S (N_S+1)} \pm (N_T+(\eta-1)N_S)$~\cite[(28)]{Gagatsos2020CovertCapacity}. The first term of~\eqref{eq:Holevo_info_cont} follows from the fact that the~\ac{TMSV} ring state is diagonal in the Fock basis, whereas the second term follows from the invariance of the von~Neumann entropy under phase shifts and the fact that for Gaussian states the von~Neumann entropy depends solely on the corresponding symplectic eigenvalues.

The first term of~\eqref{eq:Holevo_info_cont} can be lower bounded as follows,

\begin{align}
& \quad -\sum_{n_1=0}^{\infty} \sum_{n_2=0}^{\infty} p\left(n_1, n_2\right) \log p\left(n_1, n_2\right) \nonumber\\
&=	
	-\sum_{n_1=0}^{\infty} \sum_{n_2=0}^{\infty} p\left(n_1, n_2\right)\left[n_1 \log \left(\frac{N_T - \eta}{N_T-\eta+1}\right)+n_2 \log \left(\frac{\left(N_T-\eta\right) N_S}{N_T\left(N_S+1\right)}\right)\right. \nonumber \\
& \quad
	+	\left.
		\log \left(\frac{N_T-\eta}{\left(N_T-\eta+1\right) N_T\left(N_S+1\right)}\right)
	+	\log 	\left(
				{}_2F_1\left[n_1+1,n_2+1; 1, \frac{\eta}{\left(N_T-\eta +1\right) N_T} \right]
				\right)
		\right] 
		\nonumber\\
&\geq
	-\langle n_1\rangle \log \left( \frac{N_T-\eta}{N_T-\eta+1} \right)
	-\langle n_2 \rangle  \log \left( \frac{N_S(N_T-\eta)}{(N_S+1)N_T} \right)
	-\log \left( \frac{N_T-\eta}{(N_T-\eta+1)N_T(N_S+1)} \right)
	\nonumber\\
& \quad
   	+\log\left(1-\frac{\eta}{\left(N_T-\eta +1\right) N_T}\right)  \left( 1+ \langle n_1 \rangle + \langle n_2 \rangle  + \langle n_1 n_2 \rangle\right)
   	\label{eq:apply_2F1_up}\\
& \triangleq
	\mathbb{H}_\text{lb} \nonumber,
\end{align}
where $p\left(n_1, n_2\right)$ is defined in~\eqref{eq:p_n1_n2}, $\langle n_1 \rangle=\eta N_S+N_T$, $\langle n_2 \rangle=N_S$, $\langle n_1 n_2 \rangle=N_S(\eta+2\eta N_S +N_T)$, and~\eqref{eq:apply_2F1_up} follows by replacing the relevant terms with their moments and applying the bounds from Lemma~\ref{lem:2F1_ub} in Appendix~\ref{sec:F_bounds}.

\section{Round-Trip Communication with TMSV Ring States}
\label{sec:round_trip_comm_proof}
In the round-trip covert communication model depicted in Fig.~\ref{fig:round_trip_covert}, Bob's covariance matrix for the returned and idler modes in $(\hat{q}_R, \hat{p}_R, \hat{q}_I, \hat{p}_I)$ form is given by:
\begin{align}
\label{eq:Bob_cov_four_mode}
\boldsymbol{\Lambda}_{\theta, RRII}^\prime=
  \begin{bmatrix*}[l]
    B_{\text{Returned}}^\prime \matI		& C_{\text{Bob}}^\prime \mathbf{R}_\theta\\
    C_{\text{Bob}}^\prime \mathbf{R}_\theta	& S_{\text{Idler}} \matI\\
  \end{bmatrix*},
\end{align}
where { $B_{\text{Received}}^\prime \triangleq 2 \eta^\prime N_S + 2 N^\prime_T	  +1$,$C_{\text{Bob}}^\prime \triangleq 2 \sqrt{\eta^\prime}  \sqrt{N_S \left(N_S+1\right)}$},
	and { $S_{\text{Idler}} \triangleq 2 N_S+1$}. By introducing two intermediate variables, $\eta^\prime \triangleq \eta^2$ and $N^\prime_T \triangleq (1-\eta)\, \eta \, N_{B1} + (1-\eta)\,N_{B2}$, the resultant covariance matrix matches the entries of~\eqref{eq:Bob_cov_one_way}.

Since Bob's covariance matrix in the round-trip model shares the same structure as that of the one-way model, the results established in Proposition~\ref{prop:TMSV_m_penalty} remain valid under the appropriate substitution of variables.

\section{Proof of Proposition~\ref{prop:coherent_m_penalty} \\
(Difference in Holevo information between the coherent ring state and the \texorpdfstring{$m$}{m}-ary PSK coherent state)}
\label{sec:coherent_m_penalty_proof}

By applying the perturbation theorem for the von Neumann entropy~\cite[Theorem 2]{Grace2022PerturbationTheory} and letting $f(n, m \Delta)\triangleq \braket{n\vert\rho_{B}^{\theta=0}}{n-m \Delta}$ and $\lambda_{n}=p(n,\Delta=0)$ in~\eqref{eq:p_n_Delta}, we obtain

\begin{align}
\label{eq:pert-H-coherent}
\vonH{\rho_{B,m}^\textnormal{coherent}}-\vonH{{\rho}_{B}^\textnormal{coherent-ring}}
=
-
	\sum_{\Delta=1}^{\infty}
	\sum_{n=0}^{\infty}
	\abs{f(n, m\Delta)}^2 \;
	\frac{\log(\lambda_{n+m\Delta}) - \log( \lambda_n)}
    	 {\lambda_{n+m\Delta} - \lambda_n}
	+O(\norm[2]{\nu_m}^3),
\end{align}
where $\nu_m\triangleq\rho_{B,m}^\textnormal{coherent}-{\rho}_{B}^\textnormal{coherent-ring}$.

Considering the case $m=1$ and omitting the big-O term in~\eqref{eq:pert-H-coherent}, we obtain
\begin{align}
	0
&<
	\sum_{\Delta=1}^{\infty}
	\sum_{n=0}^{\infty}
	\abs{f(n, \Delta)}^2 \;
	\frac{\log(\lambda_{n+\Delta}) - \log( \lambda_n)}
    	 {\lambda_{n+\Delta} - \lambda_n}
    	 \nonumber \\
&=	
	\sum_{n=0}^{\infty}
	\sum_{\Delta=1}^{n}
	\frac{n!}{(n-\Delta)!}
	\frac{d^2 a^{2n} c^{2 \Delta}}{\Delta! \Delta!}
	{}_1 F_1\left[n+1; \Delta+1 ; z \right]^2
	\frac{\Delta \log(a) + \log
			\left(
				\frac{	{}_1 F_1\left[n+\Delta+1;1 ; z \right]}
		     		 {	{}_1 F_1\left[n+1;1 ; z        \right]}
 			\right)
 	}
 	{
 	 \left( d \; a^n \right)
 	 \left( a^\Delta \; {}_1 F_1\left[n+\Delta+1;1 ; z \right]- {}_1 F_1\left[n+1;1 ; z        \right]\right)
 	} \nonumber \\
&=	
	\sum_{n=0}^{\infty}
	\sum_{\Delta=1}^{n}
	\frac{n!}{(n-\Delta)!}
	\frac{d \; a^{n} c^{2 \Delta}}{\Delta! \Delta!}
	{}_1 F_1\left[n+1; \Delta+1 ; z \right]
	 \left(
	 \frac{\Delta \log(a) + \log
			\left(
				\frac{	{}_1 F_1\left[n+\Delta+1;1 ; z \right]}
		     		 {	{}_1 F_1\left[n+1;1 ; z        \right]}
 			\right)
 	}
 	{
 	\frac{a^\Delta \; {}_1 F_1\left[n+\Delta+1;1 ; z \right]}
 	     {{}_1 F_1\left[n+1;1 ; z        \right]}
 	- 1
 	}
	\right)
 	\nonumber \\
 &\quad \times
 	\left(
 	\frac{{}_1 F_1\left[n+1; \Delta+1 ; z \right]}
 	     {{}_1 F_1\left[n+1; 1 ; z \right]}
 	\right)
 	\nonumber \\
&\leq
	\sum_{n=0}^{\infty}
	\sum_{\Delta=1}^{n}
	\frac{n!}{(n-\Delta)!}
	\frac{d \; a^{n} c^{2 \Delta}}{\Delta! \Delta!}
	{}_1 F_1\left[n+1; \Delta+1 ; z \right]
	\left(
	\Delta^2  \left( \abs{\log a} + \log(1+z) \right)^2+2
	\right)
	\label{eq:H_perturbed_poly_coherent}
\end{align}
where~\eqref{eq:H_perturbed_poly_coherent} follows by
$
\left(
	 \frac{\Delta \log(a) +
	 		\log
			\left(
				\frac{	{}_1 F_1\left[n+\Delta+1;1 ; z \right]}
		     		 {	{}_1 F_1\left[n+1;1 ; z        \right]}
 			\right)
 	}
 	{
 	\frac{a^\Delta \; {}_1 F_1\left[n+\Delta+1;1 ; z \right]}
 	     {{}_1 F_1\left[n+1;1 ; z        \right]}
 	- 1
 	}
	\right)
<
	\Delta^2  \left( \abs{\log a} + \log(1+z) \right)^2+2
$
and
$0< \left(
 	\frac{{}_1 F_1\left[n+1; \Delta+1 ; z \right]}
 	     {{}_1 F_1\left[n+1; 1 ; z \right]}
 	\right)<1$.

\par
	Considering a general $m$ and evaluating~\eqref{eq:H_perturbed_poly_coherent}, we obtain
\begin{align}
	0
&< 
	\sum_{n=m}^{\infty}
	\sum_{\Delta=1}^{\lfloor \frac{n}{m} \rfloor}
	\frac{n!}{(n-m\Delta)!}
	\frac{d \; a^{n} c^{2 m\Delta}}{(m\Delta)! (m\Delta)!}
	{}_1 F_1\left[n+1; m\Delta+1 ; z \right]
	\left(m^2\Delta^2  \left( \abs{\log a} + \log(1+z) \right)^2+2 \right)
	\nonumber\\
&\leq 
	\sum_{n=m}^{\infty}
	\sum_{\Delta=1}^{\lfloor \frac{n}{m} \rfloor}
	\frac{n!}{(n-m\Delta)!}
	\frac{d \; a^{n} c^{2 m\Delta}}{(m\Delta)! (m\Delta)!}
	\left( 1+ \frac{z}{m\Delta+1}\right)^n
	e^z
	\left(m^2 \Delta^2  \left( \abs{\log a} + \log(1+z) \right)^2+2 \right)
	\nonumber\\
&\leq
	\frac{e^z}{N_T+1} \exp \left(-\frac{\eta \tau|\alpha|^2}{N_T}\right) \sum_{\Delta=1}^{\infty} \frac{\left(\eta \tau|\alpha|^2 / N_T^2\right)^{m \Delta}}{(m \Delta)!}\left(m^2\Delta^2 K+2\right) \frac{q_{\Delta}^{m \Delta}}{\left(1-q_{\Delta}\right)^{m \Delta+1}}
	\nonumber\\
&\triangleq
	\xi(m) \label{eq:xi_m_coherent},
\end{align}
where $K=\left(\left|\log \frac{N_T}{N_T+1}\right|+\log \left(1+\frac{\eta \tau|\alpha|^2}{N_T\left(N_T+1\right)}\right)\right)^2$, and $q_{\Delta}=\frac{N_T}{N_T+1}\left(1+\frac{\eta \tau|\alpha|^2}{N_T\left(N_T+1\right)(m \Delta+1)}\right)$. Assuming $N_T>\frac{\eta\tau\abs{\alpha}^2}{2}-1$, we have $0<q_{\Delta}<1$.
Our numerical simulations based on the Walrus library approximate $\chi_{m-\textnormal{PSK}}^\textnormal{coherent} \approx 3.47\times 10^{-5}$ and $\xi(m) \approx 2.45\times 10^{-9}$ when $m=2$, $\eta=0.5$, $\tau \abs{\alpha}^2=10^{-4}$ and $N_T=1$, indicating that the $O(\norm[2]{\nu_m}^3)$ term in~\eqref{eq:pert-H-coherent} is negligible in the parameter regime of interest.

\section{Proof of Theorem~\ref{thm:coherent_ring_Holevo_info} (Holevo information of the coherent ring state)}
\label{sec:coherent_ring_Holevo_info_proof}
We now evaluate the Holevo information

\begin{align}
\chi_{\textnormal{ring}}^\textnormal{coherent}
&= 
		\vonH{ \frac{1}{2 \pi}\int_0^{2\pi} \rho_{BI}^\theta \; d\theta}
	- 	\frac{1}{2 \pi}\int_0^{2\pi} \vonH{ \rho_{BI}^{\theta} } \; d\theta \nonumber \\
&=	\label{eq:three_mode_coherent_tensor}
	\vonH{ \frac{1}{2 \pi}\int_0^{2\pi} \rho_{B}^\theta \; d\theta \otimes \rho_I} - \vonH{ \rho_{B}^{\theta=0} \otimes \rho_I } \\
&=	\label{eq:three_mode_coherent_H_one_mode}
	\vonH{ \frac{1}{2 \pi}\int_0^{2\pi} \rho_{B}^\theta \; d\theta} - \vonH{ \rho_{B}^{\theta=0}} \\
&= \label{eq:three_mode_coherent_p_n}
	-\sum_{n=0}^\infty
	p(n) \log p(n)
	-
	g\left( \frac{\Lambda-1}{2} \right),
\end{align}
where~\eqref{eq:three_mode_coherent_tensor} follows from the independence of the modes $B$ and $I$,~\eqref{eq:three_mode_coherent_H_one_mode} follows from the fact that the entropy of an independent two-mode state equals the sum of the individual entropies with the second mode being a pure state and therefore having zero entropy, $p\left(n\right)=p\left(n, \Delta=0\right)$ defined in~\eqref{eq:p_n_Delta}, and $\Lambda = 2N_T+1$ in~\eqref{eq:three_mode_coherent_p_n}.

We lower bound the first term in~\eqref{eq:three_mode_coherent_p_n}
\begin{align}
&\quad
	-\sum_{n=0}^{\infty} p\left(n\right) \log p\left(n\right)
	\nonumber\\
&=
	-\log 	\left( 
				e^{\left(
				-\frac{\eta \tau \abs{\alpha}^2}{N_T}\right)}
				\frac{1}{N_T+1}
			\right)
	- \langle n \rangle \log	\left(
				\frac{N_T}{N_T+1}
			\right)	
	-\sum_{n=0}^\infty
			p(n) \log {}_1 F_1 \left[n+1; 1 ; \frac{\eta \tau \abs{\alpha}^2}{N_T\left(N_T+1\right)}\right]\nonumber\\
&=
	 \frac{\eta \tau \abs{\alpha}^2}{N_T}
	+\log(N_T+1) -\left(N_T+\eta\tau \abs{\alpha}^2 \right) \log N_T 
	+ \left(N_T+\eta\tau \abs{\alpha}^2 \right) \log (N_T+1) \nonumber\\
& \qquad \qquad \qquad
	-\sum_{n=0}^\infty
			p(n) \log {}_1 F_1 \left[n+1; 1 ; \frac{\eta \tau \abs{\alpha}^2}{N_T\left(N_T+1\right)}\right] \label{eq:three_mode_coherent_mean_n}\\
&\geq
	\eta \tau \abs{\alpha}^2
	  \left( \frac{1}{N_T} - \log \frac{N_T}{N_T+1} \right)
	 +g(N_T)-\left( N_T+\eta\tau\abs{\alpha}^2 \right)\log\left( 1+ \frac{\eta \tau \abs{\alpha}^2}{N_T\left(N_T+1\right)}\right)-\frac{\eta \tau \abs{\alpha}^2}{N_T\left(N_T+1\right)} \label{eq:three_mode_coherent_lb} \\
&=
	\eta \tau \abs{\alpha}^2
	  \left( \frac{1}{N_T+1} - \log \frac{N_T}{N_T+1} \right)
	 +g(N_T)-\left( N_T+\eta\tau\abs{\alpha}^2 \right)\log\left( 1+ \frac{\eta \tau \abs{\alpha}^2}{N_T\left(N_T+1\right)}\right), \nonumber
\end{align}
where~\eqref{eq:three_mode_coherent_mean_n} follows by $\langle n \rangle = N_T + \eta\tau \abs{\alpha}^2$,~\eqref{eq:three_mode_coherent_lb} follows by the ${}_1F_1$ upper bound in Lemma~\ref{lem:1F1_ub}.

Let $N_S = \tau \abs{\alpha}^2$, we have
\begin{align*}
&	\quad
	\lim_{N_S \rightarrow 0^+}
	-\sum_{n=0}^\infty
	p(n) \log p(n)
	-
	g\left( \frac{\Lambda-1}{2} \right)	\\
&=
	\eta N_S \log \left(1+\frac{1}{N_T}\right)
	-\frac{2 N_T+1}{2 N_T \left(N_T+1\right)^2} \eta ^2 N_S^2
	+O\left(N_S^3\right) \\
&=
	\eta N_S \log \left(1+\frac{1}{N_T}\right)+O\left(N_S^2\right).
\end{align*}

\section{Channel Reliability and Resolvability}
\label{sec:covert_lemma}

Building on techniques from~\cite{KhabbaziOskouei2019UnionBound,Khatri2019SecondorderCoding,Anshu2019BuildingBlocks,Wilde2017PositionbasedCoding,Wang2024ResourceefficientEntanglementassisted}, we analyze a random coding ensemble $\mathcal{C}$ and establish the reliability and resolvability bounds that will later be combined to yield a $\varepsilon$-reliable and $\delta$-covert code with a secret key $K$ available when necessary.
\begin{lemma}
\label{lem:one_way_covert_lemma}
Let $\varepsilon \in (0,1)$ and $\kappa \in \left(0, \frac{\delta}{2}\right)$. Let $W$ and $K$ denote the message and secret key, with $M$ and $L$ the cardinalities of their respective sets. Fix a quantum channel $\mathcal{G}: \rho_{X A I} \mapsto \rho_{X B I E}$ and an encoder family $\left\{\mathcal{E}^{(w,k)}_{X^nS^nI^n \rightarrow X^n A^n I^n}\right\}_{w,k}$. Let $\mathcal{C}$ be a random coding ensemble. If for sufficiently large $n$ the cardinalities satisfy
\begin{align}
	& 	\log M \geq n \mathbb{D}\left(\rho_{X B I} \| \rho_X \otimes \rho_{BI}\right)
		+ \sqrt{n \mathbb{V}\left(\rho_{X BI} \| \rho_X \otimes \rho_{BI}\right)}\Phi^{-1}(\varepsilon) + O(\log n) ,\nonumber\\
	&	\log ML
		\leq
		\max\Bigl\{	
		n \mathbb{D}\left(\rho_{X E} \| \rho_X \otimes \rho_{E}\right)
		- \sqrt{n \mathbb{V}\left(\rho_{X E} \| \rho_X \otimes \rho_{E}\right)}\Phi^{-1}\left(\frac{\kappa^2}{4}\right) + O(\log n)
		, \log M \Bigr\} ,\nonumber
\end{align}
then the ensemble satisfies the reliability and resolvability bounds
\begin{align*}
	& 	\mathbb{E}_{\mathcal{C}, K}\{\mathbb{P}(\widehat{W} \neq W \mid K)\} \leqslant \varepsilon ,\\
	& 	\mathbb{E}_{\mathcal{C}}\left\{\frac{1}{2}
		\left\|\hat{\rho}_{E}^n-\rho_{n,E}^{\otimes n}\right\|_1\right\} \leqslant \kappa ,
\end{align*}
where $\Phi(x)$ is the Gaussian~\ac{CDF}.
\end{lemma}

\section{Details for Theorem~\ref{thm:TMSV_one_way_covert} \\ (Covertness Analysis for One-Way Communication with TMSV Ring States)}
\label{sec:TMSV_one_way_covert_proof}

Upper bounding the trace distance by the fidelity $F(\rho,\rho) \triangleq \norm[1]{\sqrt{\rho} \sqrt{\rho}}$,
\begin{align*}
	\frac{1}{2}\norm[1]{{\rho}_{n,{E}}^{\otimes n}-{\rho}_{0,{E}}^{\otimes n}} 
\leq
	\sqrt{1-F \left({\rho}_{n,{E}}^{\otimes n},{\rho}_{0,{E}}^{\otimes n} \right)^2}
=
	\sqrt{1-F \left({\rho}_{n,{E}},{\rho}_{0,{E}} \right)^{2n}}.
\end{align*}
By choosing $\smash{F({\rho}_{n,{E}},{\rho}_{0,{E}})^2 \geq 1- \frac{\delta^2}{n}}$, it follows that $F({\rho}_{n,{E}},{\rho}_{0,{E}})^{2n} \geq 1- \delta^{2}$, and thus $ \sqrt{1-F({\rho}_{n,{E_1 E_2}},{\rho}_{0,{E_1 E_2}})^{2n}} \leq \delta$. Because both ${\rho}_{n,{E}}$ and ${\rho}_{0,{E}}$ are Gaussian states, we have the closed-form expression for the fidelity~\cite[(5.2)]{Marian2012UhlmannFidelity}

\begin{align}
	F 	\left( 
			{\rho}_{n,{E}}
			,
			{\rho}_{0,{E}}
			\right)^2
	&= 1-\frac{(1-\eta)^2}{4 \eta N_B (1+\eta N_B)} N_S^2 + \frac{(1-\eta)^3(1+2\eta N_B)}{8 \eta^2 N_B^2 (1+\eta N_B)^2} N_S^3 +O(N_S^4)\\
	&\geq 1-\frac{(1-\eta)^2}{4 \eta N_B (1+\eta N_B)} N_S^2,
\end{align}
and we denote $ C_2 \triangleq \frac{(1-\eta)^2}{4 \eta N_B (\eta N_B+1)}=\frac{(1-\eta)^2}{4\frac{\eta}{1-\eta}N_T \left(\frac{\eta}{1-\eta}N_T +1 \right)}$ in the main document.

\section{Details for Theorem~\ref{thm:coherent_one_way_covert} \\ (Covertness Analysis for One-Way Communication with Coherent Ring States)}
\label{sec:coherent_one_way_covert_proof}

Eve's covariance matrix is given by
{ $ \left( 2 N_T^\prime+1\right) I_2$}, where $N_T^\prime \triangleq \eta N_B$ and $N_B$ is the thermal mean photon number of the channel. The corresponding mean vector is given by
{
$
	\bar{r}_{\text{Eve}}
=
\begin{bmatrix}
	\sqrt{2} \operatorname{Re}\left\{\sqrt{1-\eta \tau} \alpha \right\}
	\\
	\sqrt{2} \operatorname{Im}\left\{\sqrt{1-\eta \tau} \alpha \right\}
\end{bmatrix}.
$
}

We obtain the following expression for Eve's Fock coefficients under the ring state scheme,
\begin{align}
	\bra{n} \rho_E \ket{n}
	=
	p_{E}(n)
	\triangleq
	e^{\left(
			-\frac{ (1-\eta) \tau \abs{\alpha}^2}{N_T^\prime}
		\right)}
	\;
		\frac{{N_T^\prime}^n}{\left(N_T^\prime+1\right)^{n+1}}
	\;
	{}_1 F_1
			\left[
				n+1; 1 ; \frac{ (1-\eta) \tau \abs{\alpha}^2}
							  {N_T^\prime\left(N_T^\prime+1\right)}
			\right] \nonumber.
\end{align}

Similar to Appendix~\ref{sec:TMSV_one_way_covert_proof}, we choose $F({\rho}_{n,{E}},{\rho}_{0,{E}})^2 \geq 1- \frac{\delta^2}{n}$ to satisfy the covertness constraint, where $\rho_{n,{E}}=\sum_{n=0}^\infty p_{E}(n) \ket{n}\bra{n}$ and $\rho_{0,{E}}=\sum_{n=0}^\infty \frac{{N_T^\prime}^n}{(N_T^\prime+1)^{n+1}} \ket{n}\bra{n}$.
We have
{
\begin{align}
	F({\rho}_{n,{E}},{\rho}_{0,{E}}) 
&= 	\sum_{n=0}^\infty \sqrt{p_{E}(n) \frac{{N_T^\prime}^n}{(N_T^\prime+1)^{n+1}}} \nonumber\\
&=	e^{-\frac{(1-\eta) \tau \abs{\alpha}^2}{2 N_T^\prime}}
	\frac{1}{N_T^\prime+1}
	\sum_{n=0}^\infty
	\left(\frac{{N_T^\prime}}{N_T^\prime+1} \right)^n
	\sqrt{{}_1 F_1
			\left[
				n+1; 1 ; \frac{ (1-\eta) \tau \abs{\alpha}^2}
							  {N_T^\prime\left(N_T^\prime+1\right)}
			\right]} \nonumber\\
&=	e^{-\frac{(1-\eta) \tau \abs{\alpha}^2}{2 N_T^\prime}}
	\frac{1}{N_T^\prime+1}
	\sum_{n=0}^\infty
	\left(\frac{{N_T^\prime}}{N_T^\prime+1} \right)^n
	\nonumber \\
&\quad \times
	\left(	1
			+\frac{(n+1)(1-\eta) \tau \abs{\alpha}^2}{2N_T^\prime(N_T^\prime+1)}
		 	+\frac{(n+1)(1-\eta)^2 \tau^2 \abs{\alpha}^4}{8{N_T^\prime}^2(N_T^\prime+1)^2}
		 	+O(\abs{\alpha}^6)
	\right) \nonumber,
\end{align}
}
and then
\begin{align}
	F({\rho}_{n,{E}},{\rho}_{0,{E}})^2 
=
	1- \frac{(1-\eta)^2 \tau^2 \abs{\alpha}^4}{4 N_T^\prime  \left(N_T^\prime +1\right)} +O(\abs{\alpha}^6)
\geq
	1- \frac{(1-\eta)^2 N_S^2}{4 N_T^\prime  \left(N_T^\prime +1\right)}
\geq
	1- \frac{\delta^2}{n} \nonumber,
\end{align}
where $N_S=\tau \abs{\alpha}^2$.

Consequently, we obtain the power constraint $N_S \leq \frac{1}{\sqrt{C_2^{\prime}}} \frac{\delta}{\sqrt{n}}$, where $ C_2^{\prime} \triangleq \frac{(1-\eta)^2}{4 N_T^\prime  \left(N_T^\prime +1\right)}=\frac{(1-\eta)^2}{4 \frac{\eta}{1-\eta} N_T  \left(\frac{\eta}{1-\eta} N_T +1\right)}$.

\section{Details for Theorem~\ref{thm:TMSV_roundtrip_covert} \\ (Covertness Analysis for Round-Trip Communication with TMSV Ring States)}
\label{sec:TMSV_roundtrip_covert_proof}
We upper bound the covertness constraint by the sum of two two auxiliary trace distances as follows,
\begin{align}
& \;
  		\frac{1}{2} 
  		\norm[1]{\hat{\rho}_{E_1 E_2}^n- {\rho}_{0,{E_1 E_2}}^{\otimes n}} 
  		\nonumber\\ 
\leq & \;
		\frac{1}{2} 
  		\norm[1]{\hat{\rho}_{E_1 E_2}^n-{\rho}_{n,{E_1 E_2}}^{\otimes n}}
  	+\frac{1}{2}
  		\norm[1]{{\rho}_{n,{E_1 E_2}}^{\otimes n}-{\rho}_{0,{E_1 E_2}}^{\otimes n}} 
  		\label{eq:four_mode_cov_split}\\ 
\leq & \;
		(\kappa)+(\delta -\kappa)
		\nonumber\\
\leq & \;
		\delta
		\nonumber,
\end{align}
where the first term in~\eqref{eq:four_mode_cov_split} is addressed via the convex-splitting technique and the second term is bounded through a fidelity-based analysis, which effectively imposes a power constraint on the transmitted signal.
\subsection{Convex Splitting}
\label{sec:concex_split}

Using the result of Lemma~\ref{lem:one_way_covert_lemma}, we apply convex splitting to the \ac{PSK} modulation scheme. Letting $\mathcal{H}_E = \mathcal{H}_{E1} \otimes \mathcal{H}_{E2}$, there exists a random codebook $\mathcal{C}$ such that $\frac{1}{2} \norm[1]{\hat{\rho}_{E_1 E_2}^n-{\rho}_{n,{E_1 E_2}}^{\otimes n}} \leq  \kappa$. We subsequently analyze the required secret key size $L$ by evaluating the quantum relative entropy $\mathbb{D}\left(\rho_{X E_1 E_2} \| \rho_X \otimes \rho_{E_1 E_2}\right)$ using the Holevo information $\chi_\textnormal{ring}^\textnormal{Eve}\triangleq \vonH{ \frac{1}{2 \pi}\int_{0}^{2\pi} \rho_{E_1 E_2}^\theta d\theta }-\frac{1}{2\pi}\int_{0}^{2\pi}\vonH{\rho_{E_1 E_2}^{\theta}} d \theta$, as detailed in Appendix~\ref{sec:round_trip_key}.

\subsection{Power Constraint}
\label{sec:four_mode_fidelity}
To bound the second term in~\eqref{eq:four_mode_cov_split}, we define the no-communication state $\rho_{0,E_1E_2}$ as the zero-mean Gaussian state with the covariance matrix given in~\eqref{eq:Eve_cov_four_mode} with $N_S = 0$ and a uniformly distributed phase $\theta$.
We define { ${\rho}_{n,{E_1 E_2}} \triangleq \sum_{k=0}^{m-1} \mathcal{E}^{\theta=\frac{2\pi k}{m}}_{S\rightarrow E_1 E_2}\;\rho_{S}$}, where $\rho_{S}$ denotes the signal arm of the \ac{TMSV} state, and $\mathcal{E}^{\theta=\frac{2\pi k}{m}}_{S\rightarrow E_1 E_2}$ represents the quantum channel mapping from system $S$ to Eve's systems $E_1$ and $E_2$, with the encoder applying a phase shift of $\theta = \frac{2\pi k}{m}$ to the signal.

We aim to characterize the state $\rho_{S}$ such that the following inequality is satisfied:
\begin{align*}
	\frac{1}{2}\norm[1]{{\rho}_{n,{E_1 E_2}}^{\otimes n}-{\rho}_{0,{E_1 E_2}}^{\otimes n}} 
	\leq
		\delta - \kappa .
\end{align*}
Based on the fidelity $F(\rho,\rho)=\norm[1]{\sqrt{\rho} \sqrt{\rho}}$, we have
\begin{align*}
			\frac{1}{2}\norm[1]{{\rho}_{n,{E_1 E_2}}^{\otimes n}-{\rho}_{0,{E_1 E_2}}^{\otimes n}} 
		&\leq
			\sqrt{1-F \left({\rho}_{n,{E_1 E_2}}^{\otimes n},{\rho}_{0,{E_1 E_2}}^{\otimes n}
			 \right)^2} \\
  		&=
  			\sqrt{1-F \left({\rho}_{n,{E_1 E_2}},{\rho}_{0,{E_1 E_2}} \right)^{2n}}.
\end{align*}
By choosing $F \left({\rho}_{n,{E_1 E_2}},{\rho}_{0,{E_1 E_2}} \right)^2 \geq 1- \frac{(\delta - \kappa)^2}{n}$, it follows that $F \left({\rho}_{n,{E_1 E_2}},{\rho}_{0,{E_1 E_2}} \right)^{2n} \geq 1- (\delta - \kappa)^{2}$, and thus $ \sqrt{1-F \left({\rho}_{n,{E_1 E_2}},{\rho}_{0,{E_1 E_2}} \right)^{2n}} \leq \delta - \kappa $, which matches the requirements.
Because the fidelity is jointly concave, we obtain
\begin{align}
	F({\rho}_{n,{E_1 E_2}},{\rho}_{0,{E_1 E_2}})
	&= F \left( 
			\int_0^{2\pi} \frac{1}{2 \pi}
			U_{\theta} \;
			{\rho}_{n,{E_1 E_2}}^{\theta=0} \;
			U_{\theta}^\dagger d \theta,
			\int_0^{2\pi} \frac{1}{2 \pi}
			U_{\theta} \;
			{\rho}_{0,{E_1 E_2}}^{\theta=0} \;
			U_{\theta}^\dagger d \theta
		\right) \\
	& \geq
		\int_0^{2\pi} \frac{1}{2 \pi} \; F 
			\left( 
			U_{\theta} \;
			{\rho}_{n,{E_1 E_2}}^{\theta=0} \;
			U_{\theta}^\dagger,
			U_{\theta} \;
			{\rho}_{0,{E_1 E_2}}^{\theta=0} \;
			U_{\theta}^\dagger 
			\right) d \theta \\
	&=	F 	\left( 
			{\rho}_{n,{E_1 E_2}}^{\theta=0}
			,
			{\rho}_{0,{E_1 E_2}}^{\theta=0}
			\right)
\end{align}
Because both ${\rho}_{n,{E_1 E_2}}^{\theta=0}$ and ${\rho}_{0,{E_1 E_2}}^{\theta=0}$ are Gaussian states, we have the closed-form expression for the fidelity~\cite[(5.2)]{Marian2012UhlmannFidelity}

\begin{align}
	F 	\left( 
			{\rho}_{n,{E_1 E_2}}^{\theta=0}
			,
			{\rho}_{0,{E_1 E_2}}^{\theta=0}
			\right)^2
	&= \frac{1}{1+ 0 \cdot N_S + C_2 N_S^2- C_3 N_S^3 + O(N_S^4)} \label{eq:four_mode_f_T1}\\
	&\geq \frac{1}{1+ C_2 N_S^2} \\
	&= 1-C_2^{\prime \prime} N_S^2+C_4^{\prime \prime} N_S^4 +O(N_S^5)
		\label{eq:four_mode_f_T2}\\
	&\geq 1-C_2^{\prime \prime} N_S^2,
\end{align}
where Taylor expansions are applied in~\eqref{eq:four_mode_f_T1} and~\eqref{eq:four_mode_f_T2}, and
\begin{align}
	C_2^{\prime \prime}=
	\biggl( 
& 	 N_{\mathrm{B} 1}^3\left(2 \eta^2 N_{\mathrm{B} 2}+\eta^2-\eta+1\right)
	+N_{\mathrm{B} 1}^2\left(4 \eta^3 N_{\mathrm{B} 2}^2+\eta\left(4 \eta^2+3\right) N_{\mathrm{B} 2}+1\right) \nonumber\\
& 	+N_{\mathrm{B} 1} N_{\mathrm{B} 2} \eta\left(2 \eta^3 N_{\mathrm{B} 2}^2+\eta\left(3 \eta^2+\eta+3\right) N_{\mathrm{B} 2}+\eta^3+2 \eta^2+\eta+2\right) \nonumber\\
& 	+\eta^2 N_{\mathrm{B} 2}^2\left(\eta N_{\mathrm{B} 2}+1\right)
	\biggr) \times
	\frac	{(1-\eta )^2}
			{N_{\text{B1}} \left(2 \eta ^2 N_{\text{B2}}+\eta ^2-\eta +1\right)+\eta  N_{\text{B2}}} \nonumber\\
& \times
	\frac{1}{4 \eta ^2 N_{\text{B1}} N_{\text{B2}} \left(N_{\text{B1}} \left(\eta ^2 N_{\text{B2}}+\eta ^2-\eta +1\right)+\eta  N_{\text{B2}}+1\right)}. \label{eq:C2_prime_prime}
\end{align}

Thus,
\begin{align*}
	F 	\left( 
			{\rho}_{n,{E_1 E_2}}^{\theta=0}
			,
			{\rho}_{0,{E_1 E_2}}^{\theta=0}
			\right)^2
\geq
	1-C_2^{\prime \prime} N_S^2					&\geq 1-\frac{(\delta -  \kappa)^2}{n}, \\
 	\frac{(\delta -  \kappa)^2}{n} 				&\geq C_2^{\prime \prime} N_S^2, \\
	\frac{\delta -  \kappa}{ \sqrt{n}} \frac{1}{\sqrt{C_2^{\prime \prime}}} &\geq N_S
\end{align*}
becomes the power constraint. The achievable rate under the $N_S$ constraint is optimal up to a constant factor.

\subsection{Secret Key Required by Convex Splitting}
\label{sec:round_trip_key}
The upper bound on the required secret key rate is given by { $\max \left\{0 , \frac{\log ML - \log L}{n} \right\}$}, where we evaluate { $\frac{\log ML}{n}$} using the Holevo information $\chi_\textnormal{ring}^\textnormal{Eve}\triangleq \vonH{ \frac{1}{2 \pi}\int_{0}^{2\pi} \rho_{E_1 E_2}^\theta d\theta }-\frac{1}{2\pi}\int_{0}^{2\pi}\vonH{\rho_{E_1 E_2}^{\theta}} d \theta$, which quantifies the information accessible to Eve from her received states.
Let
\begin{align*}
		{\rho}_{E_1E_2}^\textnormal{Eve-ring}
	&\triangleq
		 \frac{1}{2 \pi}\int_{0}^{2\pi} \rho_{E_1 E_2}^\theta d\theta\\
	&=
		\sum_{n_1=0}^\infty
		\sum_{n_2=0}^\infty
		p_{\text{Eve}}(n_1,n_2)
		\ket{n_1,n_2}
		\bra{n_1,n_2},
\end{align*}
where
\begin{align}
\label{eq:four_mode_Eve_p_n1_n2}
	p_{\text{Eve}}(n_1,n_2)= a^{n_1} \; b^{n_2} \; c \; {}_2F_1[n_1+1,n_2+1;1;z],
\end{align}
\begin{align*}
	a &= \frac{\eta ^2 N_{\text{B1}} N_{\text{B2}}+(1-\eta ) N_S \left(N_{\text{B1}}+\eta  N_{\text{B2}}\right)}{\eta ^2 N_{\text{B1}} N_{\text{B2}}+(1-\eta ) N_S \left(N_{\text{B1}}+\eta  N_{\text{B2}}\right)+(1-\eta )^2 N_{\text{B1}}+\eta  N_{\text{B2}}+\eta  (1-\eta ) N_S} ,\\
	b &= \frac{\eta ^2 N_{\text{B1}} N_{\text{B2}}+(1-\eta ) N_S \left(N_{\text{B1}}+\eta  N_{\text{B2}}\right)}{N_{\text{B1}} \left(\eta ^2 N_{\text{B2}}+\eta -\eta  N_S+N_S\right)+(1-\eta ) N_S \left(\eta  N_{\text{B2}}+1\right)} ,\\
	c &= \frac{1}{\eta ^2 N_{\text{B1}} N_{\text{B2}}+(1-\eta ) N_S \left(N_{\text{B1}}+\eta  N_{\text{B2}}\right)+(1-\eta )^2 N_{\text{B1}}+\eta  N_{\text{B2}}+\eta  (1-\eta ) N_S} \\
	  & \quad \times
	  	\frac{\eta ^2 N_{\text{B1}} N_{\text{B2}}+(1-\eta ) N_S \left(N_{\text{B1}}+\eta  N_{\text{B2}}\right)}{N_{\text{B1}} \left(\eta ^2 N_{\text{B2}}+\eta -\eta  N_S+N_S\right)+(1-\eta ) N_S \left(\eta  N_{\text{B2}}+1\right)} ,\\
	z &= \frac{1}{\eta ^2 N_{\text{B1}} N_{\text{B2}}+(1-\eta ) N_S \left(N_{\text{B1}}+\eta  N_{\text{B2}}\right)+(1-\eta )^2 N_{\text{B1}}+\eta  N_{\text{B2}}+\eta  (1-\eta ) N_S}\\
	  & \quad \times
	  	\frac{(1-\eta )^2 \eta  \left(N_{\text{B1}}-N_S\right){}^2}{\eta ^2 N_{\text{B1}} N_{\text{B2}}+(1-\eta ) N_S \left(N_{\text{B1}}+\eta  N_{\text{B2}}\right)+\eta  N_{\text{B1}}+(1-\eta ) N_S} .
\end{align*}

To simplify the expression of Eve's entropy of the mixed state, we bound the hypergeometric function $_2F_1[n_1+1,n_2+1;1;z]$ by a polynomial of $n_1$ and $n_2$.

Given $n_1,n_2 >0$ and $1>z>0$, we have
\begin{align*}
	_2F_1[n_1+1,n_2+1;1;z]
		&	= 		\sum_{k=0}^{\infty} \frac{(n_1+1)_k \; (n_2+1)_k}{(1)_k \; k!}z^k \\
		&	\geq 	\sum_{k=0}^{1} \frac{(n_1+1)_k (n_2+1)_k}{k!}z^k \\
		&	=		1+(n_1+1)(n_2+1)z.
\end{align*}
Because $\log(1+x)\geq x - \frac{1}{2}x^2$ when $x \geq 0$, we have
\begin{align}
\label{eq:F_lower_bound}
	\log \; _2F_1[n_1+1,n_2+1;1;z]
	& \geq (n_1+1)(n_2+1)z - \frac{1}{2} \left( (n_1+1)(n_2+1)z \right)^2 \nonumber \\
	& = \left( \frac{-1}{2}z^2+z \right) + \left( z-z^2 \right) \left( n_1+n_2 \right) \nonumber\\
	& \quad + \left( z-2z^2 \right) \left( n_1 n_2 \right)
			- \frac{1}{2} z^2 \left( n_1^2 + n_2^2 \right)
		\nonumber\\
	& \quad -z^2 \left( n_1^2  n_2 + n_1 n_2^2 \right)
	        - \frac{1}{2} z^2 \left( n_1^2 n_2^2 \right).
\end{align}

We bound Eve's entropy of the mixed state as
\begin{align}
&\quad
		\mathbb{H}{({\rho}_{E_1E_2}^\textnormal{Eve-ring})} \nonumber\\
&= 
		-\sum_{n_1=0}^{\infty} \sum_{n_2=0}^{\infty}
    	p_{\text{Eve}}(n_1,n_2) \log \; p_{\text{Eve}}(n_1,n_2) \nonumber \\
&= 
    	-\sum_{n_1=0}^{\infty} \sum_{n_2=0}^{\infty}
    	p_{\text{Eve}}(n_1,n_2) \log \; \left( a^{n_1} \; b^{n_2} \; c \; {}_2F_1[n_1+1,n_2+1;1;z] \right) \nonumber \\
&\leq
    	\quad	\left( -\log c + \frac{1}{2}z^2 -z \right)
	 	\quad 	+	\langle n_1 \rangle \left( - \log a \right)
	 	\quad 	+	\langle n_2 \rangle \left( - \log b \right) \nonumber \\
&	\quad 	+	\left( \langle n_1 \rangle + \langle n_2\rangle \right)
				\left( z^2-z \right) 
		 	+	\left( \langle n_1^2 \rangle + \langle n_2^2\rangle \right)
				\left( \frac{1}{2} z^2 \right) \nonumber \\
&\quad 		+	\langle n_1 n_2 \rangle \left( 2z^2-z \right) 
			+	\left( \langle n_1^2 n_2 \rangle + \langle n_1 n_2^2 \rangle \right)
				\left( z^2 \right) 	
			+	\langle n_1^2 n_2^2 \rangle \left( \frac{1}{2}z^2 \right) \nonumber\\
& \triangleq
	\mathbb{H}_{\text{upper bound}}{({\rho}_{E_1E_2}^\textnormal{Eve-ring})}.
\end{align}

Assuming $\theta=0$, Eve's covariance matrix in terms of $(\hat{q}_{E_1}, \hat{p}_{E_1}, \hat{q}_{E_2}, \hat{p}_{E_2})$ is given as
\begin{align}
\label{eq:Eve_cov_four_mode}
\boldsymbol{\Lambda}_{E_1E_1E_2E_2,\theta=0}=
  \begin{bmatrix*}[l]
    E_{\text{Forward}} 	\matI_2 	& -C_{\text{Eve}} 	\matI_2\\
    -C_{\text{Eve}}		\matI_2     & S_{\text{Backward}} \matI_2\\
  \end{bmatrix*},
\end{align}
where
\begin{align*}
	E_{\text{Forward}}	&= 2 \eta N_{B1} + 2(1-\eta)N_S+1 \\
	C_{\text{Eve}}		&= 2 (1-\eta) \sqrt{\eta}  (N_{B1}-N_S) \\
	S_{\text{Backward}}	&= 2 (1-\eta)^2 N_{B1} + 2\eta N_{B2}+ 2(1-\eta)\eta N_S+1.
\end{align*}
By Mathematica's $\text{Eigensystem}[i \Omega\cdot \boldsymbol{\Lambda}_{\theta=0, FFBB}]$, we have two symplectic eigenvalues:
\begin{align}
\label{eq:Eve_four_mode_eigenval_1}
	\Lambda_{\text{Eve},1} 
&=
	\sqrt{
		\frac{+(E_{\text{Forward}}+S_{\text{Backward}})
				\sqrt{4 C_{\text{Eve}}^2+(E_{\text{Forward}}-S_{\text{Backward}})^2}
				+2 C_{\text{Eve}}^2+E_{\text{Forward}}^2+S_{\text{Backward}}^2}
			{2}
		} \\
\label{eq:Eve_four_mode_eigenval_2}
	\Lambda_{\text{Eve},2} 
&=
	\sqrt{
		\frac{-(E_{\text{Forward}}+S_{\text{Backward}})
				\sqrt{4 C_{\text{Eve}}^2+(E_{\text{Forward}}-S_{\text{Backward}})^2}
				+2 C_{\text{Eve}}^2+E_{\text{Forward}}^2+S_{\text{Backward}}^2}
			{2}
		}
\end{align}
Then the entropy term based on the symplectic eigenvalues can be written as
\begin{align*}
			\mathbb{H}\left( {\rho}_{E_1E_2}^{\theta=0} 	\right)
			=
    	  	  g\left( \frac{\Lambda_{\text{Eve},1}-1}{2}  	\right) 
    	  	+ g\left( \frac{\Lambda_{\text{Eve},2}-1}{2}  	\right).
\end{align*}

We upper bound { $\frac{\log ML}{n}$} by $\mathbb{H}_{\text{upper bound}}({\rho}_{E_1E_2}^\textnormal{Eve-ring}) - \mathbb{H}\left( \rho_{E_1E_2}^{\theta=0} \right)$, and the final expression obtained via \textit{Mathematica} takes the form $C_1 + C_2 N_S + O(N_S^2)$. Due to the presence of the constant term  $C_1$, we conclude that the convex splitting in this scenario requires a secret key whose rate is upper bounded by an $O(1)$ scaling with respect to the number of channel uses $n$.

\section{Auxiliary Results (Bounds on Hypergeometric Functions)}
\label{sec:F_bounds}

\begin{lemma}[$_2F_1$ upper bound]
\label{lem:2F1_ub}
\begin{align*}
	{}_2F_1[ \alpha+1, \beta+1;1;z] \leq (1-z)^{-(\alpha+1)(\beta+1)},
\end{align*}
where $\alpha, \beta \in \mathbb{N}_0, z \in \mathbb{R}^*$ and $z \in(0,1)$.
\end{lemma}
\begin{proof}
\begin{align*}
	_2F_1[\alpha+1,\beta+1;1;z]
	& = \sum_{k=0}^\infty \binom{\alpha+k}{k}  \binom{\beta+k}{k} z^k  \\
	\left(\frac{1}{1-z} \right)^{(\alpha+1)(\beta+1)}
	&=  \sum_{k=0}^\infty \binom{(\alpha+1)(\beta+1)+k-1}{k} z^k 
\end{align*}

It is equivalent to prove
$f(k) \triangleq \binom{\alpha+k}{k}  \binom{\beta+k}{k} \leq \binom{(\alpha+1)(\beta+1)+k-1}{k} \triangleq g(k).$

When $k=0$, we have $f(0) = g(0) =1$.

For the ratio test, we have
\begin{align*}
  \frac{f(k+1)}{f(k)} &\leq \frac{g(k+1)}{g(k)} ,\\
  \frac{(\alpha+k+1)(\beta+k+1)}{(k+1)^2} &\leq \frac{(\alpha+1)(\beta+1)+k}{k+1} ,\\
  0  &\leq \alpha \beta k.
\end{align*}
And the last inequality is aways true.
\end{proof}

\begin{lemma}[Contiguous relations of ${}_2F_1$ modified from~\cite{Rakha2011NewContiguous}]
\label{lem:F_ratio}
    \begin{align}
    1 \leq \frac{{}_2F_1\left[ \alpha+1, \beta+1;1;z\right]}{{}_2F_1\left[ \alpha, \beta;1;z\right]} \leq  \frac{1}{(1-z)^2}\left( 2+ z \frac{\beta}{\alpha} + \frac{\alpha}{\beta} -\frac{1}{\alpha}-\frac{1}{\beta} \right),
    \end{align}
where $\alpha, \beta \in \mathbb{N}^+, z \in \mathbb{R}^*$ and $0 \leq z <1$.
\end{lemma}

\begin{lemma}[$\Delta$ relations of ${}_2F_1$]
\label{lem:F_ratio_consecutive_delta}
\begin{align*}
     \frac{{}_2F_1[ \alpha+\Delta, \beta+\Delta;1;z]}{{}_2F_1[ \alpha, \beta;1;z]} \leq \frac{1}{(1-z)^{2\Delta}} \left( 3+z+z (\alpha + \beta) \right)^\Delta ,
\end{align*}
where $\alpha,\beta, \Delta \in \mathbb{N}^+$ and $0< z <1.$

\begin{proof}
Consider the case when $\alpha \leq \beta$, we can write that
\begin{align}
	\frac{{}_2F_1[ \alpha +\Delta, \beta +\Delta;1;z]}{{}_2F_1[ \alpha, \beta;1;z]}
&\leq
	\frac{1}{(1-z)^{2\Delta}} \prod_{i=0}^{\Delta-1} \left( 2+z \; \frac{\beta+ i}{\alpha + i} + \frac{\alpha+i}{\beta+i} -\frac{1}{\beta+i} -\frac{1}{\alpha+i} \right) \label {eq:F_ratio_consecutive_delta_a} \\
& \leq
	\frac{1}{(1-z)^{2\Delta}} \prod_{i=0}^{\Delta-1}\left( 2+z \; (\frac{\beta}{\alpha + i}+\frac{i}{\alpha + i}) + \frac{\alpha+i}{\beta+i} \right) \nonumber \\
& \leq
	\frac{1}{(1-z)^{2\Delta}} \left( 2+z \; (\beta + 1) + 1 \right)^\Delta \nonumber \\
& \leq
	\frac{1}{(1-z)^{2\Delta}} \left( 3+z+z (\alpha + \beta) \right)^\Delta \nonumber
	,
\end{align}
where~\eqref{eq:F_ratio_consecutive_delta_a} follows by Lemma~\ref{lem:F_ratio}.

Consider the other case when $\alpha > \beta$, we can write that
\begin{align}
	\frac{{}_2F_1[ \alpha +\Delta, \beta +\Delta;1;z]}{{}_2F_1[ \alpha, \beta;1;z]}
&\leq
	\frac{1}{(1-z)^{2\Delta}} \prod_{i=0}^{\Delta-1} \left( 2+z \; \frac{\alpha+ i}{\beta + i} + \frac{\beta+i}{\alpha+i} -\frac{1}{\alpha+i} -\frac{1}{\beta+i} \right) \label{eq:F_ratio_consecutive_delta_b} \\
&\leq
	\frac{1}{(1-z)^{2\Delta}} \prod_{i=0}^{\Delta-1} \left( 2+z \; (\frac{\alpha}{\beta + i}+\frac{\ i}{\beta + i}) + \frac{\beta+i}{\alpha+i} \right) \nonumber \\
&\leq
	\frac{1}{(1-z)^{2\Delta}} \left( 2+z \; (\alpha + 1) + 1 \right)^\Delta \nonumber \\
&\leq
	\frac{1}{(1-z)^{2\Delta}} \left( 3+z+z (\alpha + \beta) \right)^\Delta \nonumber
	,
\end{align}
where~\eqref{eq:F_ratio_consecutive_delta_b} follow by Lemma~\ref{lem:F_ratio}.
\end{proof}
\end{lemma}


\begin{lemma}[$_1F_1$ upper bound]
\label{lem:1F1_ub}
\begin{align*}	
	{}_1F_1[n+1,1;z] \leq {}_1F_1[1,1;z] \; (1+z)^n = e^z\; (1+z)^n 
\end{align*}
for $n\in \mathbb{N}^+$, $z\in \mathbb{R}$ and $0<z<1$.
\begin{proof}
\begin{align}	
	\frac{{}_1 F_1[n+1 , 1 ; z]}{{}_1 F_1[1 , 1 ; z]}
&=
	\prod_{i=1}^n \frac{{}_1 F_1[i+1 ; 1 ; z]}{{}_1 F_1[i ; 1 ; z]} \nonumber\\
&\leq
	\prod_{i=1}^n(1+z) \label{eq:1F1_cont}\\
&=
	(1+z)^n \nonumber,
\end{align}
where~\eqref{eq:1F1_cont} follows from~\cite[(13.4.4)]{Abramowitz1972HandbookMathematical} together with the monotonicity of the confluent hypergeometric function. 
\end{proof}
\end{lemma}

\bibliographystyle{quantum}
\bibliography{references_Quantum.bib}
\end{document}